\documentclass[twocolumn,twoside]{revtex4}

\usepackage[utf8]{inputenc} 
\usepackage[T1]{fontenc}

\usepackage{pdfpages}
\usepackage{graphicx,graphics}
\usepackage{amsmath,epsf,subfigure}
\usepackage{color}
\usepackage{amsfonts,amsmath}
\usepackage{bm,subfigure}
\usepackage{epsf}
\usepackage{amssymb}
\usepackage{afterpage}
\usepackage{float}
\usepackage{chemmacros} 
\usepackage{tikz-cd}
\usepackage{MnSymbol}
\usepackage{comment}
\usepackage{tikz}
\usepackage{chemfig}
\usepackage{soul}

\newcommand{\e}{\epsilon}

\newcommand{\Tc}{T_{\rm cycle}}
\newcommand{\Ti}{T_{\rm inhib}}
\newcommand{\dd}{\ddagger}

\newcommand{\beq}{\begin{equation}}
\newcommand{\eeq}{\end{equation}}
\newcommand{\harp}{\xrightleftharpoons}
\newcommand{\xa}{\xrightarrow[]}

\newcommand{\s}{\sigma}

\usepackage{amsmath}

\begin{document}

\title{Design principles, growth laws, and competition of minimal autocatalysts}

\author{Yann Sakref}
\author{Olivier Rivoire}

\affiliation{Gulliver, CNRS, ESPCI, Universit\'e PSL, Paris, France.}

\begin{abstract}

The apparent difficulty of designing simple autocatalysts that grow exponentially in the absence of enzymes, external drives or ingenious internal mechanisms severely constrains scenarios for the emergence of evolution by natural selection in chemical and physical systems. Here, we systematically analyze these difficulties in the context of one of the simplest and most generic autocatalysts: a dimeric molecule that duplicates by templated ligation.  We show that despite its simplicity, such an autocatalyst can achieve exponential growth autonomously. This only requires that the rate of the spontaneous dimerization, the interactions between molecules, and the concentrations of substrates and products are in appropriate ranges. We also show, however, that it is possible to design as simple sub-exponential autocatalysts that have an advantage over exponential autocatalysts when competing for a common resource. We reach these conclusions by developing a general theoretical framework based on kinetic barrier diagrams. Besides challenging commonly accepted assumptions in the field of the origin of life, our results provide a blueprint for the experimental realization of elementary autocatalysts exhibiting a form of natural selection, whether on a molecular or colloidal scale.

\end{abstract}

\maketitle

The path from simple chemical systems to complex living organisms is believed to hinge on a pivotal point at which one molecule, or a set of molecules, gain the capability to catalyze their own formation, hence constituting an autocatalytic system~\cite{orgel_molecular_1992, eigen_steps_1996, luisi_emergence_2006, zachar_new_2010, hanopolskyi_autocatalysis_2021}. When several such systems are formed from a common molecule, the faster ones hinder the growth of the slower ones, which can lead to their exclusion. This elementary form of natural selection is thought to set the stage for Darwinian evolution~\cite{orgel_molecular_1992, eigen_steps_1996, luisi_emergence_2006}. Mathematically, exclusion occurs whenever replicators grow exponentially using a common limiting resource, in which case only the fastest growing replicator can possibly survive~\cite{szathmary_sub-exponential_1989, scheuring_survival_2001, kiedrowski_selection_2005}.

Molecular replication in extant living organisms relies on enzymatic catalysis and involves a large network of coupled reactions. Non-enzymatic autocatalysts have been designed in a variety of artificial systems and at a variety of scales, from the molecular and colloidal scale up to the macroscopic scale~\cite{von_kiedrowski_self-replicating_1986, zielinski_autocatalytic_1987,tjivikua_self-replicating_1990, wang_self-replication_1997,lee_self-replicating_1996, issac_approaching_2002, kamioka_autocatalysis_2010, colomb-delsuc_exponential_2015,kosikova_two_2019,zhuo_litters_2019}. At the molecular scale, the simplest systems implement template replication, where the formation of a new complex $AB$ from its constituents $A$ and $B$ is catalyzed by a previously formed complex $AB$. However, such non-enzymatic molecular autocatalysts are generally found to exhibit sub-exponential growth, where the number $x$ of autocatalysts follows the phenomenological equation \(dx/dt = kx^n\) with \(n < 1\), associated with polynomial growth, \(x(t) \sim t^{1/(1-n)}\)~\cite{dugas_minimal_1993, hanopolskyi_autocatalysis_2021}. A growth order of \(n \approx 1/2\) is typically observed, also known as parabolic growth due to the relationship \(x(t) \sim t^2\)~\cite{von_kiedrowski_self-replicating_1986, sievers_self-replication_1998, dugas_minimal_1993, hanopolskyi_autocatalysis_2021}. Von Kiedrowski showed that this sub-exponential growth originates from product inhibition, the propensity of autocatalytic templates to inhibit their catalytic activity by binding to each other~\cite{dugas_minimal_1993}. Sub-exponential autocatalysts are not mutually exclusive and are therefore often regarded as implementing only a primitive and weak form of selection, if they are taken into account at all in the emergence of natural selection~\cite{dugas_minimal_1993, szathmary_replicators_1997, scheuring_survival_2001, colomb-delsuc_exponential_2015}.

For this reason, much work has been devoted to the design of autocatalysts that alleviate product inhibition. The first type of solutions involve external drives applied in a cyclical pattern, such as heat~\cite{braun_pcr_2004, wang_self-replication_2011, zhuo_litters_2019}, mechanical stress~\cite{carnall_mechanosensitive_2010}, light~\cite{zhang_engineering_2007, zhang_accelerated_2014}, tidal cycling~\cite{fernando_stochastic_2007}, or magnetic fields~\cite{dempster_self-replication_2015}. Approaches based on the intrinsic properties of the autocatalyst rather than external factors have also been proposed~\cite{penrose_self-reproducing_1957, penrose_mechanics_1958, penrose_self-reproducing_1959, virgo_evolvable_2012, abe_destabilizing_2004, osuna_galvez_traceless_2019, zhang_engineering_2007,cabello-garcia_handhold-mediated_2021, juritz_minimal_2022}. For instance, at the molecular level, the affinity between autocatalysts can be diminished by coupling the formation of a bound within autocatalysts to the breaking of a bound between autocatalysts~\cite{abe_destabilizing_2004, osuna_galvez_traceless_2019}, or by entropic mechanisms like toeholds and handholds strand displacements when using nucleic acids~\cite{zhang_engineering_2007,cabello-garcia_handhold-mediated_2021, juritz_minimal_2022}. These approaches, although effective in specific settings, raise several questions. First, it is often ambiguous whether the proposed mechanism mitigates product inhibition, the binding of two preformed autocatalysts, or accelerate product release, the unbinding of a newly formed autocatalyst from a preformed catalyst, which is known to impact the growth rate but not the growth order of autocatalysis~\cite{virgo_evolvable_2012, cabello-garcia_handhold-mediated_2021, zhang_engineering_2007, michaelis_amplification_2014}. Second, these mechanisms are often idiosyncratic to the context in which they are designed, which limits the scope of their applicability. Finally, and perhaps most importantly, the designs are generally intricate and fine-tuned, which defeats the purpose of studying autocatalysis as a means to understand the accretion of complexity.

This poses a major problem for origin-of-life scenarios based on autocatalysis. 
Consequently, most scenarios currently focus on autocatalytic networks composed of multiple molecules rather than single-molecule autocatalysts~\cite{kim_cross-catalytic_2004, lincoln_self-sustained_2009, zeravcic_self-replicating_2014, semenov_autocatalytic_2016, cafferty_robustness_2019, blokhuis_universal_2020, ameta_darwinian_2021}. This orientation reflects the belief that these networks are more likely to emerge spontaneously~\cite{kauffman_cellular_1971, eigen_principle_1977, eigen_steps_1996, szostak_optimal_2011, liu_mathematical_2018, hordijk_history_2019, hordijk_molecular_2019}. However, networks raise similar challenges~\cite{sievers_self-replication_1998, hordijk_influence_2018}, as well as posing new ones, e.g., the likely appearance of parasites~\cite{eigen_selforganization_1971, matsumura_transient_2016}.

In either case, whether based on a single species or a network of species, the design of autocatalysts has so far remained in the realm of empirical studies. In particular, no theoretical argument rules out the possibility of simple autocatalysts growing exponentially or establishes the minimum requirements that they must satisfy. Here, we propose to fill this gap by showing through a systematic approach that simple and generic autocatalysts are designable, although with limitations that we clarify. By simple, we mean autocatalysts composed of very few elements (two) with no internal structure or internal degree of freedom. By generic, we mean an entropic mechanism of autocatalysis by proximity that is present in any chemistry or colloidal system subject to thermal noise. Our approach is to treat autocatalysis as a special case of catalysis -- namely when the product is a catalyst -- and to apply a previously developed methodology to define, construct and optimize minimal catalysts~\cite{munoz-basagoiti_computational_2023,sakref_kinetic_2023}. However, this is only a starting point: as we show, this methodology needs to be extended to account for the constraints arising from the identity between products and catalysts, which introduces a fundamental distinction between catalysis and autocatalysis. As a result, we demonstrate that it is possible to design simple generic autocatalysts that grow exponentially, but that it is equally possible to design simple sub-exponential autocatalysts that out-compete them in conditions of resource limitation.

\section{Methods}

\subsection*{Model}

\begin{figure}[t]
   \centering
   \includegraphics[width=1\linewidth]{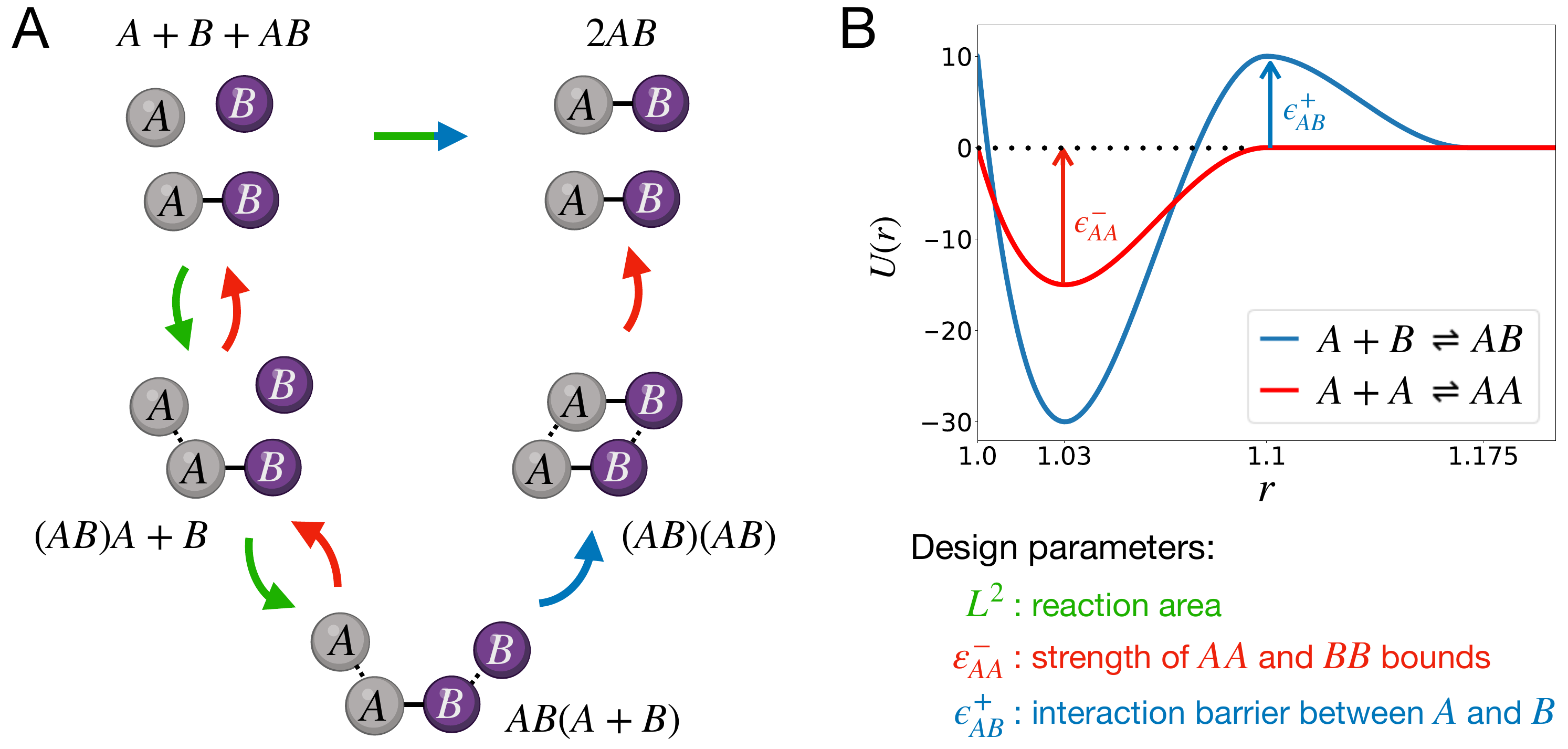}
   \caption{Model for the design of minimal autocatalysts \textbf{A.}~Autocatalytic cycle in which particles $A$ and $B$ can attach to dimer $AB$ catalyzing their dimerization. The scheme represents $A$ binding $AB$ before $B$ but the reverse order is also possible. Green arrows indicate diffusion processes, dependent on the area $L^2$. Red arrows represent dissociations of two identical particles, dependent on the interaction strength $\e_{AA}^-$. Blue arrows represent association between distinct particles, dependent on the interaction barrier $\e_{AB}^+$. The spontaneous reaction involves both diffusion and association and is indicated by a two-colored arrow. \textbf{B.}~Potentials by which particles interact (Materials and methods). Between identical particles, the potential depth is $\e_{AA}^-$, and association is diffusion-limited. Between distinct particles, the potential depth is very large (infinite), and association limited by a barrier $\e_{AB}^+$. \label{fig: Figure1}}
\end{figure}

We study the design of autocatalysts $AB$ composed of two units $A$ and $B$ which catalyze their own formation through a templating reaction summarized by $AB+A+B\to 2AB$ (Fig.~\ref{fig: Figure1}A). Guided by simplicity, we consider for $A$ and $B$ spherical particles of same diameter $\s$, immersed in a thermal bath at temperature $T$ within a two-dimensional box of dimension $L\times L$. For illustration and to indicate the experimentally feasibility of our design, we take inspiration from DNA-coated colloids~\cite{rogers_using_2016} and present numerical results using a short-range, pairwise potential with a reverse barrier (Materials and methods). As represented in Fig.~\ref{fig: Figure1}B, this potential features a cutoff distance of $r_c = 1.1\ \sigma$, and a minimum at $r_{\rm min} = 1.03\ \sigma$~\cite{wang_lennard-jones_2020, cui_comprehensive_2022}. Thus, only two parameters are left to specify the interaction between two particles of types $X$ and $Y$: the energy barrier for dimer association, $\e_{XY}^+$, and the energy barrier for dimer dissociation, $\e_{XY}^-$. 

With two particle types, $A$ and $B$, we generally need six parameters to specify the interaction potentials. We reduce this number to two by making additional simplifying assumptions. First, we consider that the dimerization reaction, $A+B\to AB$, is irreversible ($\e_{AB}^-=\infty$), and that the interaction between $A$ and $B$ is therefore described by a single parameter $\e_{AB}^+$, the association barrier. Second, we consider that the interaction potentials between two $A$ or two $B$ are the same, with same depth ($\e_{AA}^-=\e_{BB}^-$) and no association barrier ($\e_{AA}^+=\e_{BB}^+=0$), leaving a single parameter $\e_{AA}^-$, the interaction strength, to describe their interaction. Also to simplify the analysis, we assume that no molecule of size larger than four can be formed. As summarized in Fig.~\ref{fig: Figure1}C, the model has a total of three dimensionless parameters: $L/\sigma$, $\e_{AB}^+/k_BT$ and $\e_{AA}^-/k_BT$ where $k_B$ is the Boltzmann constant. Without loss of generality, we set $\sigma = 1$ to define the length scale, and $k_BT = 1$ to define the energy scale. To these three physical parameters, we must add the current concentrations of molecular species. Again for simplicity, we assume that $A$ and $B$ have the same concentration $[A]=[B]$. The only remaining parameter is then $[AB]$, the concentration of free products, or $[AB]_{\rm tot}$, the total concentration of products, including those in complex with other species. 

\subsection*{Questions}

In the context of this model, the questions raised in the introduction can be formulated as follow: What are the physical parameters $L$, $\e_{AB}^+$, $\e_{AA}^-$ and the chemical conditions $[A]$ and $[AB]$ for (i)~optimal autocatalysis, that is, leading to a maximal acceleration of the dimerization reaction $A+B\to AB$ by a pre-existing $AB$? (ii)~exponential growth, $d[AB]_{\rm tot}/dt=k[AB]_{\rm tot}$? (iii)~exclusion of an alternative autocatalyst $AD$ sharing with $AB$ a common constituent $A$?

\subsection*{Approach}

As an intermediate step towards the design of an autocatalyst $AB$, we first consider a dimeric catalyst $C=A'B'$, which is distinguishable from $AB$, but has identical physical properties ($\e_{A'B'}^+=\e^+_{AB}$ and $\e^-_{A'A'}=\e^-_{B'B'}=\e^-_{AA}$). Studying the catalysis $C+A+B\to C+AB$ enables us to apply and extend the methods previously developed to design a minimal catalyst for the reverse reaction, the dissociation of $AB$ into $A+B$~\cite{munoz-basagoiti_computational_2023}, and provides a basis for subsequently exposing the nuances between catalysis and autocatalysis. 

More precisely, we derive constraints on the design of minimal autocatalysts in four steps, starting from standard catalysis in the simplest setting and progressively introducing elements of feedback inherent to autocatalysis: (1)~We determine the conditions under which a dimer $C=A'B'$ can accelerate the dimerization reaction $A+B\to AB$. This is done by comparing the time for the spontaneous formation of a dimer $AB$ in the presence and in the absence of a $C$~\cite{sakref_kinetic_2023}. (2) Next, given a catalyst $C=A'B'$, we determine the conditions for its optimal efficiency. This is done by minimizing the cycling time $\Tc^0$, defined as the mean time taken by one $C=A'B'$ to turn a substrate $A+B$ into a product $AB$. 

Following previous work~\cite{munoz-basagoiti_computational_2023}, we solve (1) and (2) in conditions that are most favorable for catalysis, namely in the absence of any product $AB$~\cite{sakref_kinetic_2023}. (3) One unique feature of autocatalysis, however, is that it necessarily takes place in the presence of products, since the catalyst is itself a product. Products generally cause product inhibition, whereby a product binds a catalyst and inhibits its activity. We first analyze the consequence of product inhibition in standard catalysis, when the catalyst $C=A'B'$ differs from the product $AB$ and show that it increases the mean cycling time to $\Tc=\Tc^0+\Ti$ with an additional time $\Ti$ that depends on the concentration $[AB]$ of products. (4) Finally, we apply the previous results to $C=AB$ and highlight how autocatalysis departs from catalysis. In particular, while for standard catalysis the rate of product formation is, when assuming the spontaneous reaction to be negligible, proportional to the concentration of catalysts, i.e., of the form $d[AB]_{\rm tot}/dt=k[C]$ with $k=1/\Tc$, this is no longer the case for autocatalysis because $\Tc$ depends on $[AB]$ with $AB=C$.

\begin{figure*}[!t]
   \centering
   \includegraphics[width=.9\linewidth]{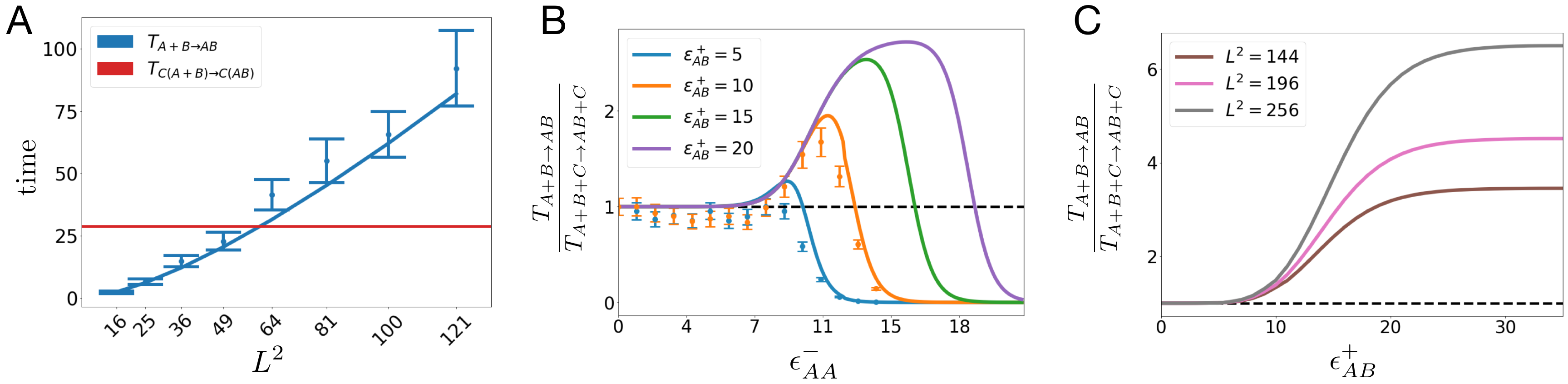}
   \caption{Conditions for catalysis of the dimerization $A+B\to AB$. \textbf{A}. Mean times for the dimerization $A+B \to AB$ in the absence of $C=A'B'$ (in blue) and for the dimerization $C(A+B) \to C(AB)$ when $A,B$ are kept attached to $C$ (in red). A necessary condition for catalysis is $T_{A+B\to AB}>T_{C(A+B)\to C(AB)}$~\cite{sakref_kinetic_2023}. Since the first time scales with the reaction area $L^2$ while the second is independent of it, catalysis requires a sufficiently large value of $L^2$. The lines are from the Markov model presented in the text and the bars from MD. \textbf{B.} The catalytic efficiency of $C$ shows a maximum at an intermediary value of the interaction strength $\epsilon_{AA}^-$, consistent with Sabatier principle. The value of this maximum increases with the interaction barrier $\epsilon_{AB}^+$. \textbf{C.} The catalytic efficiency for optimal $\e_{AA}^-$ increases both with the reaction barrier $\e_{AB}^+$ and with the reaction area $L^2$.} \label{fig: Figure2}
\end{figure*}

\section{Design principles for minimal autocatalysts (auto)catalysts}

\subsection*{Conditions for catalysis}

To determine the conditions under which a dimer $C=A'B'$ can cause the acceleration of the reaction $A+B\to AB$, we first consider a closed system with only one particle $A$ and one particle $B$ and determine the mean time $T_{A+B\to AB}$ for a dimer $AB$ to form~\cite{munoz-basagoiti_computational_2023}. We compare this time to $T_{C+A+B\to C+AB}$, the mean time for $AB$ to form when a prospective catalyst $C$ is added. Catalysis occurs when this later time is shorter than the former, that is, when the relative catalytic efficiency defined by the ratio $T_{A+B\to AB}/T_{C+A+B\to C+AB}$ is superior to $1$. In previous work, we showed that a molecule acts as an (auto)catalyst in presence of multiple molecules $A$ and $B$ only if it acts as one in presence of a single $A$ and a single $B$~\cite{sakref_kinetic_2023}. 

A first necessary condition is for the dimerization onto the catalyst to be faster than the spontaneous reaction in the bulk, i.e., $T_{C(A+B)\to C(AB)}<T_{A+B\to AB}$~\cite{sakref_kinetic_2023}. As expected from Arrhenius equation, we verify with molecular dynamics (MD) simulations that both these times scale exponentially with the association barrier $\e_{AB}^+$ when it is sufficiently large (Fig.~S1): $T_{A+B\to AB} \approx L^2e^{\e_{AB}^+}$ and $T_{C(A{+}B)\to CAB} \approx e^{\e_{AB}^+}$. Catalysis therefore requires a minimal area $L^2$ to occur. For the design at hand, we find that an area of $(L/\s)^2 \gtrsim 50$ is necessary (Fig.~\ref{fig: Figure2}A). 

Assuming such sufficiently large area $L^2$, we next study the impact of the two physical parameters, $\e_{AB}^+$ and $\e_{AA}^-$. To extend this study beyond the range of parameter values accessible by MD, we approximate the dynamics by a Markov model with five distinct states, corresponding to the various states of bonding between the autocatalyst, the monomers $A$ and $B$, and the product $AB$ (Fig.~\ref{fig: Figure1}A). Formally, the catalytic cycle is described by
\beq\label{eq:kin}
C+A+B\harp[\rho_{-1}]{\rho_1} CA+B\harp[\rho_{-2}]{\rho_2} C(A{+}B)\xa{\rho_3} C(AB)\xa{\rho_4} C+AB,
\eeq
closed by adding $C+A+B\harp[\rho_{-0}]{\rho_0}C+AB$ to represent the spontaneous reaction without any interaction with the catalyst. Here we assume that $A$ and $B$ are equivalent and therefore do not differentiate between $CA+B$ and $CB+A$. We also assume that release occurs in a single step, which is a good approximation when $\e_{AA}^-$ is sufficiently large (Fig~S2). We take the dependence of the rate on the parameters to be given by
\begin{equation}\label{eq:cycle}
    \begin{aligned}
        \rho_{1} \approx  2L^{-2},& \quad \rho_{2} \approx L^{-2}, \quad \rho_{3} \approx e^{-\e_{AB}^+}, \quad \rho_4 \approx e^{-2\e_{AA}^-},\\
        &\rho_{-1} \approx e^{-\e_{AA}^-},\quad \rho_{-2} \approx 2e^{-\e_{AA}^-}.
    \end{aligned}
\end{equation}
Pre-factors can be introduced to obtain a better fit to the MD simulations (Supplementary Materials and Fig.~S3), but they have no major impact on the results (Fig.~S8) and are omitted here to simplify the presentation.

The catalytic efficiency depends both on the interaction strength $\e_{AA}^-$ and on the association barrier $\e_{AB}^+$. For a given association barrier $\e_{AB}^+$, we observe an optimal interaction strength $\e_{AA}^-$ (Fig.~\ref{fig: Figure2}B). This observation follows Sabatier's principle, which applies broadly to catalytic systems with no internal degrees of freedom~\cite{medford2015sabatier, rivoire_how_2023}, and states that an optimal interaction between a catalyst and its substrate must strike a balance between too weak an interaction that cannot hold the substrates until they react, and too strong an interaction that cannot release the product rapidly. 

A second observation is that larger association barriers $\e_{AB}^+$ enable greater relative catalytic efficiencies $T_{A+B\to AB}/T_{C+A+B\to C+AB}$ (Fig.~\ref{fig: Figure2}B) This is again a generic feature: the larger the barrier for the spontaneous reaction, the more potential for catalysis. In fact no catalysis can occur if the barrier is too small~\cite{munoz-basagoiti_computational_2023}. Finally, increasing the reaction area also increases the relative efficiency of the catalyst (Fig.~\ref{fig: Figure2}C). This is simply the consequence of increasing the mean time of the spontaneous dimerization reaction in solution without changing  the dimerization reaction on the catalyst. 

In summary, catalysis of the reaction $A+B\to AB$ is favored by a large reaction barrier $\e_{AB}^+$, a large reaction volume $L^2$ and a particular, finite value of the interaction strength $\e_{AA}^-$ that depends on $\e_{AB}^+$ and $L^2$.

\subsection*{Optimal cycling time in the absence of products}

\begin{figure*}[!t]
   \centering
   \includegraphics[width=0.95\linewidth]{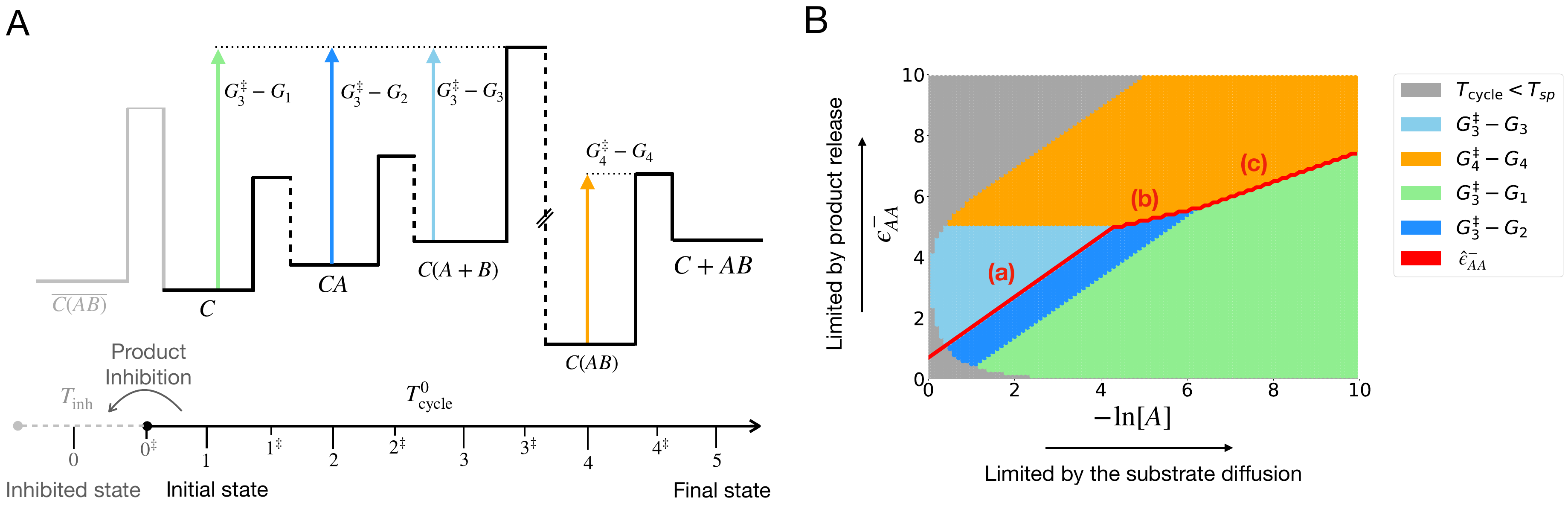}
   \caption{Kinetic energy diagram and limiting barriers -- \textbf{A.} Kinetic energy diagram associated with the Markov chain described by Eq.~\eqref{eq:kin} and Eq.~\eqref{eq:bar}. Local minima represent states while local maxima represent transition states, at levels corresponding to the rates between successive states, as given by Eq.~\eqref{eq:rates}. The mean cycling time is approximated by the largest difference between successive levels, as indicated in Eq.~\eqref{eq:span}. In this illustration, it is given by $G^\ddagger_3-G_1$ (light green) but other values of the parameters can lead to other limiting barriers. As in Fig.~\ref{fig: Figure1}, different colors refer to different processes: light blue for dimerization, darker blue for association of a substrate followed by dimerization, green for diffusion of both substrates followed by dimerization, and orange for the dissociation of the product. Backward direct barriers are indicated with dashed lines. In the presence of products, an additional state $i=0$ can be reached, representing a non-productive complex $\overline{C(AB)}$, here placed on the left of the diagram. \textbf{B.} Limiting barrier as a function of the substrate concentration $[A]$ and the interaction strength $\epsilon_{AA}^-$, for $\epsilon_{AB}^+ = 10$, and no product, $[AB] = 0$. The red line represents the optimal interaction strength with three different regimes, (a), (b), (c), depending on which two barriers are in trade-off (see Supplementary Material).} \label{fig: Figure3}
\end{figure*}

Having determined the conditions under which a molecule $C$ acts as a catalyst, we now analyze how the catalytic turnover depends on the concentration of substrates $[A]=[B]$. To this end, we can ignore the spontaneous reaction. As a first step, we also assume that products are systematically removed so that $[AB]=0$. The rates of the elementary processes along the cycle are formally obtained by replacing $L^{-2}$ by $[A]$ in Eq.~\eqref{eq:cycle}, to account for the possible presence of multiple substrates. The Markov chain for the complete cycle can be represented graphically as a kinetic energy diagram~\cite{penocchio_kinetic_2023} (Fig.~\ref{fig: Figure3}A). In this diagram, each of the five states $i$ ($i=1$ for $C$, $i=2$ for $CA$, $i=3$ for $C(A{+}B)$, $i=4$ for $C(AB)$ and $i=5$ for $C+AB$) is represented at an energy level $G_i$ and successive states are separated by transition states at energy level $G_i^\ddagger$, such that the differences of energies between states and transition states report the rates as
\begin{equation}\label{eq:rates}
    \rho_i = e^{-(G_i^\ddagger -G_i)}, \quad \rho_{-i} = e^{-(G_i^\ddagger -G_{i+1})}.\end{equation}
In this representation, the mean cycling time has a simple expression~\cite{kozuch_what_2011, park_describing_2022, penocchio_kinetic_2023},
\beq \label{eq: MFPT}
\Tc^{(0)}([A])=\sum_{1 \leq i \leq j \leq 4} e^{G_j^{\ddagger}-G_i}
\eeq
where the sum is over each pair $i \leq j$ of transition state $j$
following a ground state $i$ and where the subscript $(0)$ indicates that no product is present. This sum is typically dominated by its largest term so that 
\beq\label{eq:span}
\Tc^{(0)}([A])\approx e^{\max_{1 \leq i \leq j \leq 4}(G_j^{\ddagger}-G_i)}.
\eeq
The exponent defines the limiting barrier, also known as the energy span~\cite{kozuch_what_2011, kozuch_how_2011}, which is represented in kinetic barrier diagrams by the largest difference of energy between successive -- but not necessarily consecutive -- levels. This limiting barrier formalizes the intuitive but problematic notion of ``limiting step'', which takes only into account successive levels~\cite{kozuch_how_2011}. As we show below, reducing the estimation of the mean cycling time to the determination of the limiting barrier simplifies the analysis and the interpretation of the results without changing qualitatively the conclusions.

Limiting barriers can be of two types, direct barriers between successive states and indirect barriers between non-successive states. Direct barriers report the mean time to perform one elementary transition. The dependence of the direct barriers $G_i^\dd-G_i=-\ln\rho_i$ on the parameters is given by Eq.~\eqref{eq:cycle},
\begin{equation}\label{eq:dirb}
    \begin{aligned}
G_1^\dd-G_1&\approx  -\ln [A] -\ln 2,\\
G_2^\dd-G_2&\approx -\ln [A],\\
G_3^\dd-G_3&\approx \e_{AB}^+,\\
G_4^\dd-G_4&\approx 2\e_{AA}^-.
    \end{aligned}
\end{equation}
The first two barriers describe the diffusion of substrates to the catalyst, the third barrier the dimerization reaction on the catalyst, and the last the release of the product.

The total cycling time is, however, more than the addition of these elementary transition times. Indeed, once a state has been reached, the next elementary transition may be a backward transition and not a forward one, corresponding to a recrossing event. This is the origin of the indirect barriers between non-consecutive states, given by
\begin{equation}  \label{eq:revb}
    \begin{aligned}
G_2^\dd-G_1=&\ln\frac{\rho_{-1}}{\rho_1\rho_2}\approx-\e_{AA}^--2\ln [A] -\ln2,\\
G_3^\dd-G_1=&\ln\frac{\rho_{-1}\rho_{-2}}{\rho_1\rho_2\rho_3}\approx-2\e_{AA}^-+\e_{AB}^+-2\ln [A],\\
G_3^\dd-G_2=&\ln\frac{\rho_{-2}}{\rho_2\rho_3}\approx-\e_{AA}^-+\e_{AB}^+-\ln [A]+\ln2.
    \end{aligned}
\end{equation}
These indirect barriers are all the smaller than the backward direct barriers -- the direct barriers for the catalysis of the reverse reaction $AB\to A+B$ -- are higher. Hence, a short cycling time requires not only low forward direct barriers but also high backward direct barriers. As apparent in Eqs.~\eqref{eq:dirb} and Eqs.~\eqref{eq:revb}, the different barriers are not independent but controlled by the same physical and chemical parameters. These relationships capture the essential trade-offs involved in the design of catalysis.

For instance, at low substrate concentration, the optimal interaction energy is $\hat \e_{AA}^- = (\e_{AB}^+ - 2\ln[A])/4$. Consistent with Sabatier principle, this optimum strikes a balance between the indirect barrier for substrate binding and dimerization $G_3^\dd-G_1$ (in green in Fig.~\ref{fig: Figure3}B), which is diminished by increasing $\e_{AA}^-$, and the direct barrier for product release $G_4^\dd-G_4$ (in orange in Fig.~\ref{fig: Figure3}B), which is conversely diminished by increasing $\e_{AA}^-$. The same reasoning applies at higher substrate concentrations, where the direct barrier for product release is in trade-off with other indirect barriers related to substrate binding (segments (a) and (b) in Fig.~\ref{fig: Figure3}B, see Supplementary Material). 

The analysis of limiting barriers in kinetic barrier diagrams thus reveals how different trade-offs control the design of optimal catalysts, depending on chemical and physical parameters (Fig.~\ref{fig: Figure3}B).

\subsection*{Optimal cycling time in the presence of products}

We now extend the analysis to the presence of free products, $[AB]\neq 0$. The presence of product generally increases the mean cycling time, because a catalyst can bind to a product instead of a substrate, thus forming a non-productive complex that we denote $\overline{C(AB)}$. This non-productive complex $\overline{C(AB)}$ is physically indistinguishable from the unreleased complex $C(AB)$ that constitutes the last step along a catalytic cycle (Fig.~\ref{fig: Figure1}A) but the recognition that they are two different kinetic states is key to our analysis. Since $\overline{C(AB)}$ is a complex with a previously free $AB$, while $C(AB)$ is a complex with a newly made $AB$, they are indeed associated with two distinct constraints on catalysis, namely product inhibition and product release. Formally, Eq.~\eqref{eq:kin} already accounts for product release, and additionally accounting for product inhibition is done by extending it to include
\beq\label{eq:bar}
\overline{C(AB)} \harp[\rho_{-I}]{\rho_I} C+AB
\eeq
where $\rho_{I}\approx \rho_{-4} \approx [AB]$ and $\rho_{-I}\approx e^{-2\e_{AA}^-}$.

The total mean cycling time $ \Tc([A], [AB])$ is then increased by the mean time $\Ti([A], [AB])$ spent in the inhibited state $\overline{C(AB)}$, 
\beq\label{eq: Formula T_{cycle}}
    \Tc([A], [AB]) = \Tc^{(0)}([A]) + \Ti([A], [AB]).
\eeq
The slowdown due to product inhibition is a particular form of competitive inhibition where the product itself acts as the inhibitor~\cite{bezerra_enzyme_2016}. 

$\Ti([A], [AB])$ can be expressed by extending the kinetic barrier diagram to include a state $i=0$ associated with $\overline{C(AB)}$, leading to
\beq \label{eq: Formula T_{inhibition}}
    \Ti([A], [AB]) = \sum_{i=1}^{3} e^{G_i^\ddagger - G_0},
\eeq
where the new kinetic barriers to consider are obtained from the previous ones as
\beq
G_i^\dd-G_0=G_i^\dd-G_1+\ln\frac{\rho_{I}}{\rho_{-I}}
\eeq
for $i=1,2,3$, leading to
\begin{equation}  \label{eq: wp}
    \begin{aligned}
        G_1^\dd-G_0&\approx 2\e_{AA}^--\ln [A] + \ln [AB] -\ln 2,\\
        G_2^\dd-G_0&\approx \e_{AA}^--2\ln [A] + \ln [AB]  -\ln 2,\\
        G_3^\dd-G_0&\approx \e_{AB}^+ -2\ln [A] +  \ln [AB].
    \end{aligned}
\end{equation}

As shown in Fig.~\ref{fig: Figure4}A, those additional barriers can dominate the others, leading the mean cycling time to be limited by product inhibition, $T_{\rm cycle} \approx \Ti$. In particular, this happens for large relative concentration of products, $[AB] \gg [A]$, such that the catalyst is more likely to bind a product than a substrate, and for large interaction strength with respect to the concentration of product, $\epsilon_{AA}^-\gg-\ln [AB]/2$, such that the time spent in the inhibited complex $\overline{C(AB)}$ is long (SM). Since the barriers associated with product inhibition increase with $\e_{AA}^-$, one consequence of the accumulation of products is generally a decreased optimal interaction strength, as illustrated in Fig.~\ref{fig: Figure4}A.

\section{Growth laws for minimal autocatalysts}

\subsection*{Production rate} 

Assuming a buffered concentration of free substrates $A$ and $B$, and a negligible spontaneous reaction, the rate of product formation is obtained from the mean cycling time as~\cite{ninio_alternative_1987}
\begin{equation}
 \label{eq: main growth}
    \frac{d[AB]_{\rm tot}}{dt} = \frac{1}{T_{\rm cycle}([A], [AB])} [C]_{\rm tot},
\end{equation}
where $[AB]_{\rm tot}$ is the total concentration of products, including those which, after being formed, bind to a catalyst or a substrate, and where $[C]_{\rm tot}$ is the total concentration of catalysts, either free or bound. With standard catalysis, $[C]_{\rm tot}$ remains constant and the rate of product formation is simply proportional to it. With autocatalysis, however, $C=AB$, and the total concentration of catalysts increases as more products are formed. Eq.~\eqref{eq: main growth} becomes
\begin{equation}
    \frac{d[AB]_{\rm tot}}{dt} = \frac{1}{T_{\rm cycle}([A], [AB])} [AB]_{\rm tot},
\end{equation}
which is generally a non-linear function of $[AB]_{\rm tot}$ since $[AB]$ is itself a function of $[AB]_{\rm tot}$. Special conditions are therefore required for exponential growth to occur, where $d[AB]_{\rm tot}/dt=k [AB]_{\rm tot}$ with a rate $k$ independent of $[AB]_{\rm tot}$. 

\begin{figure*}[!t]
   \centering
   \includegraphics[width=.95\linewidth]{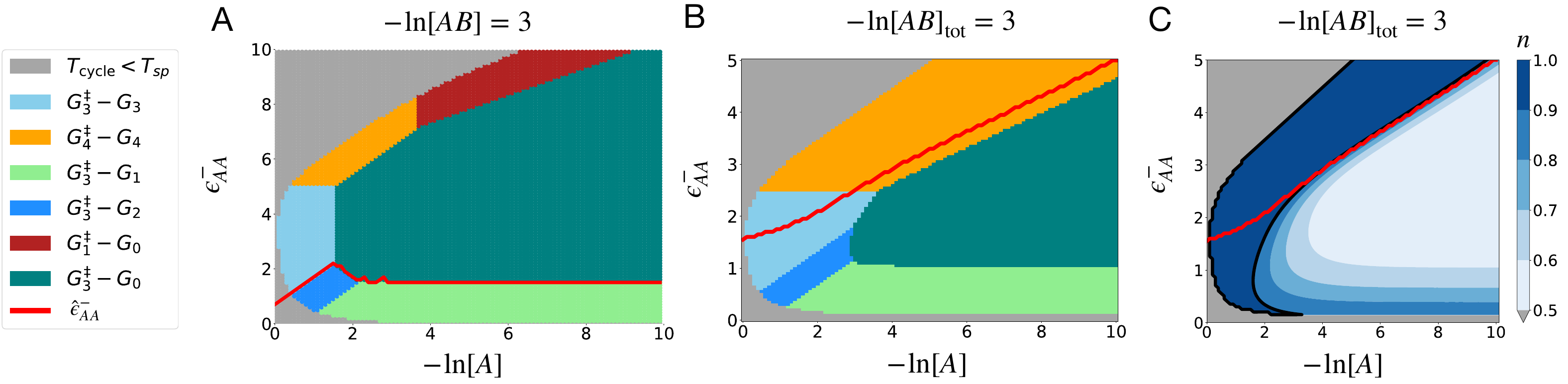}
   \caption{Limiting barriers in the presence of products. \textbf{A.} Limiting barriers for a given concentration of free product $[AB]$. Compared to Fig.~\ref{fig: Figure3}B, the limiting barrier can be associated with product inhibition (regimes in darker colors), in which case $T_{\rm cycle} \approx \Ti$. As a consequence, the optimal interaction strength $\e_{AA}^-$ is changed (red line). \textbf{B.} Limiting barriers when fixing the total concentration of autocatalyst $[AB]_{\rm tot}$ instead of the concentration of free autocatalyst $[AB]$. The results are similar at low $\e_{AA}^-$, when $[AB]_{\rm tot}\simeq [AB]$, but different at large $\e_{AA}^-$, when $[AB]_{\rm tot}\simeq [(AB)(AB)]$. In these two cases, $[\overline{(AB)(AB)}]\ll[AB]_{\rm tot}$, implying two opposite limits with no product inhibition and, therefore, exponential growth. \textbf{C.}~Reaction order $n$, as computed from simulations of the ordinary differential equations describing the Markov model (Materials and methods). In comparison to B, we see that $n<1$ even in regions where the limiting barrier is not associated with product inhibition. This is because product inhibition is always present, even when it does not control the limiting barrier. A value $n>0.9$ is nevertheless observed for a large range of parameter values (dark blue). }\label{fig: Figure4}
\end{figure*}

\subsection*{Conditions for exponential growth}

The decomposition of the cycling time in Eq.~\eqref{eq: Formula T_{cycle}} makes explicit the conditions for exponential growth to occur: since $\Ti([A], [AB])$ depends on $[AB]$ but $\Tc^{(0)}([A])$ does not, we must have $\Tc^{(0)}([A]) \gg \Ti([A], [AB])$, i.e., product inhibition must be negligible. 

Fig.~\ref{fig: Figure4}A shows that this occurs when release rates significantly exceed diffusion rates, \(\epsilon_{AA}^- \ll -\ln[A]\), which a systematic analysis of limiting barriers confirms (Supplementary Material). Fig.~\ref{fig: Figure4}A is drawn for a fixed concentration of free product $[AB]$, but it is often more informative to fix the total concentration of products, $[AB]_{\rm tot}$, which better reflects the progression of the dynamics and the consumption of resources -- a determining factor when considering competitions as below. 

When considering a fixed $[AB]_{\rm tot}$, Fig.~\ref{fig: Figure4}B also shows that product inhibition is negligible when \(\epsilon_{AA}^- \ll -\ln[A]\). This coincides with the results of Fig.~\ref{fig: Figure4}A because in this case most products are free, i.e., $[AB]_{\rm tot}\approx[AB]$, which implies that inhibiting complexes $\overline{(AB)(AB)}$ are negligible (Fig.~S6). However, a significant difference appears in the opposite limit $\e_{AA}^-\gg-\ln[A]$ where, in contrast to Fig.~\ref{fig: Figure4}A, Fig.~\ref{fig: Figure4}B shows an extended regime where product inhibition is negligible (in orange). In this regime, most products are in the form of unreleased complexes $(AB)(AB)$ and therefore also not forming inhibiting complexes $\overline{(AB)(AB)}$ (Fig.~S6). The distinction made in Eq.~\eqref{eq:bar} and Fig.~\ref{fig: Figure3}A between the physically identical but kinetically distinct states of unreleased and inhibiting complexes is critical to understand this regime. 

Reducing the analysis to the identification of limiting barriers is an approximation that provides necessary but not sufficient conditions for strictly exponential autocatalysis: a barrier associated with product inhibition may indeed contribute significantly to the cycling time even if it is not the limiting barrier. To go beyond this approximation, we approximate the dynamics with the phenomenological equation $d[AB]_{\rm tot}/dt=k[AB]_{\rm tot}^n$ (Materials and Methods) and analyze the conditions under which $n\approx 1$. We verify that these conditions are more demanding than those for which the limiting barrier is not associated with product inhibition, but nevertheless verify that autocatalytic growth is nearly exponential growth for a large number of parameters, even taking into account the constraint that the growth rate must exceed the spontaneous reaction rate (Fig.~\ref{fig: Figure4}C). 

The conditions for exponential growth, either weak or large interaction strengths, are in direct contrast to the conditions for minimal cycling time which, following Sabatier principle, requires an intermediate interaction strength (Fig.~\ref{fig: Figure2}B). As illustrated in Fig.~\ref{fig: Figure4}C, this translates into a generic trade-off between the reaction constant $k$ and the reaction order $n$. The strength of this trade-off depends, however, on the values of the interaction barrier $\e_{AB}^+$: increasing $\e_{AB}^+$ mitigates this trade-off, drawing the optimal values of $k$ and $n$ closer together (Fig.~S7). This occurs for large values of both $\e_{AB}^+$ and $\e_{AA}^-$, when $\e_{AB}^+ > \e_{AA}^- > -\ln[AB]/2$. Indeed, as \(\epsilon_{AB}^+\) increases, the optimal \(\epsilon_{AA}^-\) also increases according to Sabatier's principle, until a point where it saturates and where no free autocatalyst remains, thereby preventing product inhibition. 

\section{Competition rules for minimal autocatalysts}

\begin{figure}[!t]
   \centering
   \includegraphics[width=0.85\linewidth]{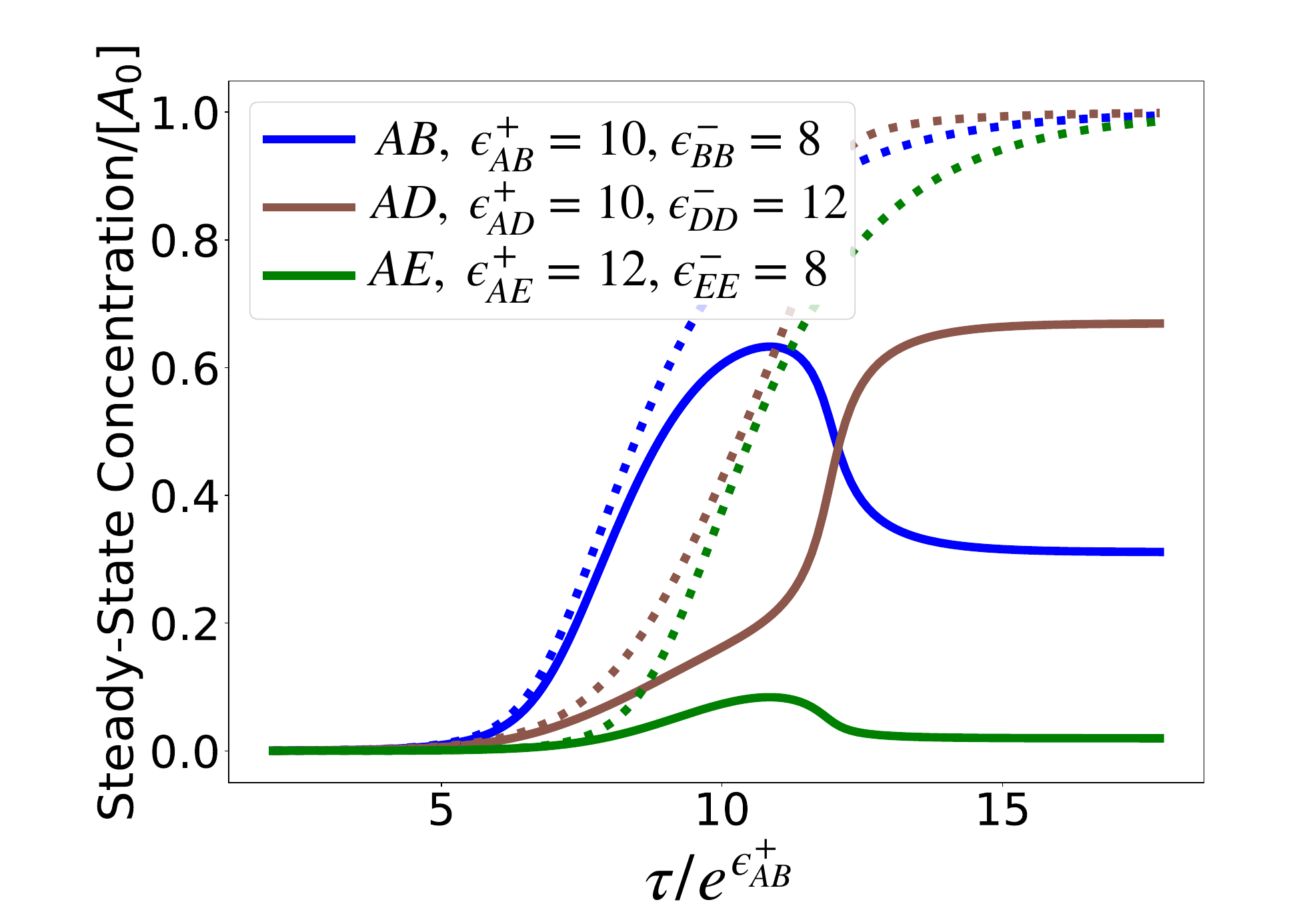}
    \caption{Competition for a common limiting resource. Steady-state concentrations of non-competing (dotted lines) and competing (plain lines) autocatalysts $AB$, $AD$, and $AE$ in a chemostat, as a function of the residence time $\tau$. The steady-state concentrations are normalized by the concentration at which the substrates are supplied, $[A]_0 = [B]_0 = [C]_0 = [D]_0 = e^{-10}$. While low residence times favor $AB$, higher residence times favor autocatalysts $AD$. The figure also illustrates how an autocatalyst of a lower efficiency, here $AE$, can be excluded.}     \label{fig: Figure5}
\end{figure}

One consequence of product inhibition is that the cycling time alone does not determine the outcome of competitions between autocatalysts. To demonstrate this, we consider in Fig.~\ref{fig: Figure5} a simple setting with three autocatalysts in a chemostat, $AB$, $AD$, and $AE$, all competing for a common resource $A$. Substrates $A$, $B$, $D$, and $E$ are supplied at a uniform constant rate $\tau^{-1}$, and all molecules are diluted at the same rate $\tau^{-1}$, so that  $\tau$ represents the typical residence time in the chemostat. 

Previous theoretical investigations have emphasized a fundamental difference between exponential ($n=1$) and sub-exponential ($n<1$) autocatalysts in such conditions: while exponential autocatalysts invariably compete to exclude one another, sub-exponential autocatalysts typically coexist~\cite{szathmary_sub-exponential_1989, scheuring_survival_2001, lifson_coexistence_2001, kiedrowski_selection_2005}. In recent work, we considered the competition of autocatalysts of different order $n$ and noted that, somewhat counterintuitively, a sub-exponential autocatalyst ($n<1$) can exclude an exponential one ($n=1$) if its reaction constant $k$ is sufficiently large~\cite{sakref_exclusion_2024}. This occurs, notably, at high dilution rates, when the mean residence time of the molecules in the chemostat is short relative to the mean cycling time, or, equivalently, when resources are scarce. In such conditions, autocatalysts are kept at a low concentration, mitigating product inhibition and making reaction constants $k$ the determining factor. This is verified in our model when competing $AB$ with $AD$, an autocatalyst of higher $n$ but lower $k$: $AD$ dominates $AB$ only for sufficiently large values of $\tau$ (Fig.~\ref{fig: Figure5}A). Thus, not only does an optimal cycling rate not guarantee dominance, but no intrinsic property of the autocatalyst guarantees it independently of the extrinsic conditions in which the competition takes place. Finally, this figure also illustrates how multiple autocatalysts competing for the same resource may either coexist or exclude each other, despite no strict exponential growth ($n$ is never strictly $1$).

\section{Discussion}

Our analysis of minimal autocatalysis reveals that, contrary to what previous attempts might have suggested, exponential autocatalysts can be designed without recourse to complex internal mechanisms, complicated geometries, or external drives. Past limitations are well-known to originate from product inhibition, the propensity of autocatalysts to bind to each other after they have been produced. Our analysis concurs in recognizing product inhibition as a fundamental constraint, but indicates that this constraint can be circumvented by an appropriate choice of parameters. Our approach also clarifies how the limitations due to product inhibition differ from those due to product release, which impacts the reaction rate but not necessarily the reaction order.

The focus on exponential growth stems from the exclusion principle that it implies, which is often considered as a core principle of natural selection~\cite{dugas_minimal_1993, scheuring_survival_2001}: two exponential autocatalysts cannot coexist if they depend on the same resource. Our results underline that exponentiality is not an intrinsic property of an autocatalyst, but crucially depends on extrinsic conditions, and that exclusion can occur in the absence of exponential growth. However, it can also be argued that the absence of strict exclusion is in itself conducive to the emergence of diversity and of evolution by natural selection~\cite{szathmary_sub-exponential_1989, zachar_new_2010}.

The key feature of our model is its definition based solely on physical principles: all possible molecules and reactions are derived from interaction potentials between elementary ``atoms''. This reveals how different rate constants are in trade-off because they depend on the same physical parameters. Our analysis of simple competitions between autocatalysts can thus go beyond previous studies where product inhibition is phenomenologically described by a reaction constant $k$ and a reaction order $n$~\cite{szathmary_sub-exponential_1989, lifson_coexistence_2001, scheuring_survival_2001}. In particular, our model demonstrates how $n$ and $k$ can be in trade-off: maximizing $n$ to achieve exponential growth ($n=1$) typically comes at the expense of a low $k$. In competitions between autocatalysts, whether a large $k$ or a large $n$ is advantageous depends on the chemical environment. If resources are abundant, autocatalysts with higher reaction order tend to prevail, but, if resources are scarce, autocatalysts with higher reaction constants have an advantage, irrespectively of their reaction order. An exclusive focus on the reaction order $n$ may therefore be misleading. 

We defined a generic and simple model with a view to its implementation in various molecular or colloidal systems. First, we chose the catalytic mechanism to be of the most basic form: the (auto)catalyst catalyzes a dimerization reaction simply by increasing the frequency of interaction between substrates when they are attached to it. This form of catalysis by proximity is universal and applies irrespective of whether the dimerization barrier is entropic or enthalpic. The parameters in our model also have their direct counterpart in almost all chemical contexts. For example, in the realm of nucleic acids, inter-dimer interactions correspond to base pairing via hydrogen bonds, while stronger intra-dimer interactions with an association barrier correspond to nearly irreversible endothermic phosphodiester covalent bonds~\cite{monnard_taming_2016, dickson_determination_2000}. In the realm of colloids whose interactions are mediated by the hybridization of complementary DNA strands or by magnetic forces, association barriers can correspond to electrostatic repulsion, to an entropic barrier due to steric effects, or to linkage-mediated interactions~\cite{dempster_self-replication_2015,angioletti-uberti_mobile_2014, rogers_mean-field_2020, wang_colloids_2012, feng_specificity_2013, zhang_multivalent_2018, kress_colloidal_2020, mcmullen_self-assembly_2022}. In this context, interaction strengths are typically of the order of a few $k_BT$ and unbinding occur within $<1$ min~\cite{rogers_kinetics_2013, cui_comprehensive_2022}. In this case, exponential growth would require interaction strengths of the order of $10$ $k_BT$, depending on the relative concentration of substrate over product (Fig.~S8). Exponential autocatalysts would then replicate within hours, in the range of experimentally accessible timescales.

For the sake of simplicity, we assumed that no molecule larger than four in size can form. This is straightforwardly the case with molecular systems that are intrinsically anisotropic~\cite{von_kiedrowski_self-replicating_1986, zhuo_litters_2019, kress_colloidal_2020} but, may be more difficult to impose on isotropic colloids~\cite{angioletti-uberti_mobile_2014}. However, a simple extension of the model translates this assumption into a constraint on the valence of atoms that is easier to implement. Our analysis indeed applies without change to the cross-catalysis of two dimers $AB$ and $A'B'$, where each type of atom is constrained to interact with at most two atoms of two different types, $A$ with $B$ and $A'$, $B$ with $A$ and $B'$, $A'$ with $A$ and $B'$, $B'$ with $A'$ and $B'$. DNA or RNA replication works by such cross-catalysis between complementary strands~\cite{sievers_self-replication_1998, zhang_engineering_2007, zhuo_litters_2019}. With spherical colloids, cross-catalysis can for instance be implemented by limiting interactions to patches~\cite{kress_colloidal_2020} (Fig.~S4). However, we constrained the size of the molecules only to simplify the analysis, and the possibility of forming larger molecules is obviously of interest on its own. 

The trade-offs that constrain our model fundamentally stem from its deliberate simplicity. In particular, the tension between chemical acceleration, on the one hand, and product release and inhibition, on the other, which underlies Sabatier principle and plays a key role in our analysis, can be overcome by a variety of mechanisms~\cite{perez-ramirez_strategies_2019, rivoire_how_2023}. In all practical cases, however, these mechanisms involve large and complex molecules. Our analysis shows that they are not prerequisites for exponential growth, or selection by exclusion. This resolves an apparent paradox in origin-of-life scenarios that seek to explain complexity as a consequence of Darwinian evolution, but require complex mechanisms for such evolution to take place.

\section*{Acknowledgments}

We acknowledge funding from ANR-22-CE06-0037. 



\section*{MATERIALS AND METHODS}

\subsection*{Molecular dynamics simulations}

Brownian molecular dynamics (MD) simulations were carried out in HoomD 3.5.0~\cite{anderson_hoomd-blue_2020}, using a time step $\Delta t=10^{-5}$, periodic boundary conditions, and a damping constant $\gamma = 10$, corresponding to a translational diffusion coefficient $k_BT / \gamma = 0.1 \text{ length}^2 / \text{time}$, comparable to values measured in experiments with colloids~\cite{zhang_multivalent_2018}. 
The potential between two particles $X$ and $Y$ is taken to be
\begin{equation}
U_{XY}(r) = 
\begin{dcases}
\epsilon_{XY}^-u\left( r\right) + \epsilon_{XY}^+ & \text{if } r \leq r_{\mathrm{c},} \\
-\epsilon_{XY}^-u\left(r - r_{\mathrm{c}}+ r_{\min}\right) & \text{if } r_{\mathrm{c}} \leq r \leq 2r_{\mathrm{c}}-r_{\min }\nonumber
\end{dcases}
\end{equation}\\
where $\epsilon_{XY}^-$ and $\epsilon_{XY}^+$ represent the activation barriers for dissociation and association, respectively. The potential $u(r)$ is a generalization of the Wang-Frenkel potential~\cite{wang_lennard-jones_2020}, with a cutoff value of $r_c = 1.1$, 
\begin{equation}
u(r) =
\begin{dcases}
\alpha\left(r_{\mathrm{c}}\right)\left[\left(\frac{\sigma}{r}\right)^2-1\right]\left[\left(\frac{r_{\mathrm{c}}}{r}\right)^2-1\right]^2 & \text{for } r \leq r_{\mathrm{c}} \\
0 & \text{for } r > r_{\mathrm{c}}.\nonumber
\end{dcases}
\end{equation}

\subsection*{Markov model}

We approximate catalytic cycles by Markov chains. With two atoms $A$ and $B$, the Markov chain involves a total of 15 transitions. First, the spontaneous reaction $A+B \to AB$, with rate $[A][B]e^{-\e_{AB}^+}$. Second, $7$ association reactions, $AB + A \to AAB$, $AB + B \to ABB$, $AAB + B \to AABB$, $ABB + A \to AABB$, $AB + AB \to \overline{ABAB}$, $A+A \to AA$ and $B+B \to BB$, with rates proportional to the reactant concentrations. Third, the dimerization reaction on the autocatalyst, $AABB \to ABAB$, with rate $[AABB]e^{-\e_{AB}^+}$. Finally, $8$ dissociation reactions, $AA \to A+A$, $BB \to B+B$, $AAB \to A + AB$, $ABB \to B + AB$, $AABB \to B + AAB$, $AABB \to A + ABB$, with rates $e^{-\e_{AA}^-}$, and $ABAB \to AB + AB$, $\overline{ABAB} \to AB + AB$ with rates $e^{-2\e_{AA}^-}$. When considering two competing autocatalysts $AB$ and $AC$ sharing a common monomer, we ignore for simplicity the complexes that they may form, of the type $(AB)(AC)$, which are less stable than homotetramers $(AB)(AB)$. 

We determine the steady state of the Markov chain by integrating numerically the ordinary differential equations that describe its dynamical evolution. We consider the system to be either closed (Fig.~\ref{fig: Figure4}C) or in a chemostat (Fig.~\ref{fig: Figure5}). In this later case, differential equations include the description of the introduction of substrates, $\emptyset \to A$ and $\emptyset \to B$, with rate $[A]_0/\tau$, and the dilution of all species, $X\to \emptyset$, with rate $1/\tau$.

\subsection*{Reaction order and reaction constant}

We estimate a reaction order $n$ and a reaction constant $k$ such that
$d[AB]_{\rm tot}/dt = k[AB]_{\rm tot}^{n}$ approximatively holds by integrating numerically the dynamical equations of the Markov chain with constant values of $[A]=[B]$, starting with $[AB]=0$ and ending when reaching the targeted value of $[AB]_{\rm tot}$. The values of $k$ and $n$ are then obtained by linear regression of $\ln(d[AB]_{\rm tot}/dt)$ against $\ln([AB]_{\rm tot})$.

\end{document}


\title{{\bf Supplementary Material}\\
\ \\
{\rm Design principles, growth laws, and competition\\ of minimal autocatalysts}\\
}
\author{Yann Sakref \& Olivier Rivoire}

\date{}

\maketitle

\tableofcontents

\newpage

\section{Model}

Our model consists of Brownian particles with isotropic or anisotropic interactions. The model is general, but setting our parameter values, we choose them to describe experiments with DNA-coated colloids, which are a possible experimental realization of our model. To capture all relevant physical trade-offs of (auto)catalysis, we fit the (auto)catalysis mediated by these particles by a Markov model calibrated to reproduce the results of molecular dynamics (MD) simulations. In this section, we detail the definition of the model, the MD simulations, and the construction of the Markov model. 

\subsection{Physical model}

We consider spherical particles of diameter $\sigma$ subjected to Brownian dynamics. We fix $\sigma = 1$ for the length scale of the system, $k_BT = 1$ for the energy scale, and take $\gamma = 10$ for the damping constant, which results in a translational diffusion coefficient $D = k_BT / \gamma = 0.1 \text{ length}^2 / \text{time}$, comparable to values measured experimentally with colloids~\cite{zhang_multivalent_2018}. 

The potential by which the particles interact is a generalization of the isotropic pairwise Wang-Frenkel potential~\cite{wang_lennard-jones_2020}, with a cutoff value of $r_c = 1.1$, represented by
\begin{equation}
\phi(r) =\begin{cases}
 \alpha\left(r_{\mathrm{c}}\right)\left[\left(\frac{\sigma}{r}\right)^2-1\right]\left[\left(\frac{r_{\mathrm{c}}}{r}\right)^2-1\right]^2 &  \text { for } r \leq r_{\mathrm{c}} \\
    0              &  \text { for } r>r_{\mathrm{c}}
\end{cases},
\end{equation}
where
\begin{equation}
\alpha\left(r_{\mathrm{c}}\right)=2\left(\frac{r_{\mathrm{c}}}{\sigma}\right)^2\left(\frac{3}{2\left(\left(\frac{r_{\mathrm{c}}}{\sigma}\right)^2-1\right)}\right)^3
\qquad{\rm and}\qquad
r_{\min }\left(r_{\mathrm{c}}\right)=r_{\mathrm{c}}\left(\frac{3}{1+2\left(\frac{r_{\mathrm{c}}}{\sigma}\right)^2}\right)^{1 / 2}.
\end{equation}
This potential has proven to be a good model for interactions involving DNA-coated colloids~\cite{cui_comprehensive_2022}. We introduce a reaction barrier by adding a mirrored reflection of this potential along the $\e_{AA}^-$-axis, with a maximum at $r=1.1\ \sigma$, and a cutoff at $r_c = 1.17\ \sigma$. 

Formally, the potential between two particles $\e_{AA}^-$ and $Y$ is taken to be
\begin{equation} \label{eq: potential MD}
U_{XY}(r) =\begin{cases} \epsilon_{XY}^-\phi\left( r\right) + \epsilon_{XY}^+ & r \leq r_{\mathrm{c}} \\ -\epsilon_{XY}^-\phi\left(r - r_{\mathrm{c}}+ r_{\min}\right) & r_{\mathrm{c}} \leq r \leq 2r_{\mathrm{c}}-r_{\min }\end{cases}
\end{equation}
where $\epsilon_{XY}^-$ and $\epsilon_{XY}^+$ represent the activation barriers for dissociation and association, respectively (Fig.~1B).

\subsection{Molecular dynamics simulations}

Molecular dynamics simulations were carried out in HoomD 3.5.0~\cite{anderson_hoomd-blue_2020}, using a time step $\Delta t=10^{-5}$ and periodic boundary conditions.

When considering anisotropic particles, we use the rigid bodies simulations implemented in HoomD~\cite{nguyen_rigid_2011, glaser_pressure_2020}. A rigid body is composed of a large inert particle with a diameter of \(\sigma\), to which smaller particles, each with a diameter of \(0.1\sigma\), are arranged on its surface. These smaller particles define patches that interact with each other via the potential given in Eq.~(\ref{eq: potential MD}) where $r_{\rm min} = 0.03 \sigma$, $r_c = 0.1$ if $\epsilon_{XY}^+ =0$, and $r_{\rm max} = 0.1$ and $r_{c} = 0.117$ otherwise. These parameters are such that the equilibrium distance between two large particles remains $1.03\ \sigma$.

\subsection{Markov models} \label{sec:model}

\begin{figure}[t]
    \centering
    \includegraphics[width=.95\linewidth]{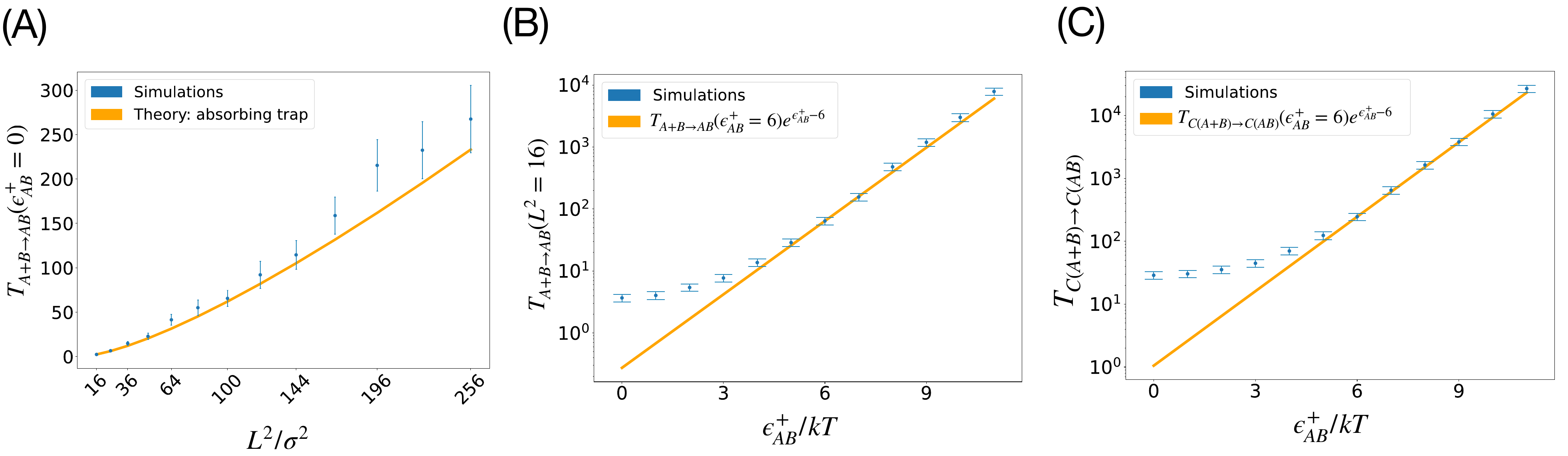} 
    \caption{\small Dimerization reactions in the bulk and on the (auto)catalyst. \textbf{(A)} In blue, molecular dynamics (MD) simulation results of the time required for two particles, initially placed randomly within a reaction area $L^2$, to diffuse toward each other. The orange line is the theoretically calculated first passage time of a particle confined within a disk of radius $L/2$ reaching an absorbing trap of radius $\sigma/2$ at the center of the disk~\cite{vot_first-encounter_2022}. \textbf{(B)} Mean dimerization time in the bulk $T_{A+B \to AB}$ as a function of the energy barrier $\e_{AB}^+$. We verify Arrhenius law for sufficiently large $\e_{AB}^+$, specifically at the value $\e_{AB}^+\gtrsim 6$ $k_BT$. \textbf{(C)} Mean dimerization time on the autocatalyst $T_{AB(A+B) \to AB(AB)}$, as a function of the energy barrier $\e_{AB}^+$, which verifies again Arrhenius law for $\e_{AB}^+ \gtrsim 6$ $k_BT$.} \label{fig:S1}
\end{figure}

In this section, we present how a Markov model is built to reproduce the dynamics observed in MD simulations. It is obtained by analyzing each step of the cycle shown in Fig.~1A of the main text, from binding to release, and applies to both an autocatalyst $AB$ and a catalyst $C = A'B'$, hereinafter collectively referred to as $C$ for the sake of generality.

First, we approximate the mean dimerization time in the absence of catalyst by Arrhenius equation~\cite{arrhenius_uber_1889} as $T_{A+B \to AB} \approx T_{d}e^{\e_{AB}^+}$, where $T_{d}$ refers to the average time needed for two particles randomly placed in the reaction area $L^2$ to come in contact. As shown in Fig.~S\ref{fig:S1}A, this time $T_d$ is well approximated by the mean first-passage time taken by a point-like particle confined within a disk of radius $R_{d} = L \pi^{-1/2}$ to reach an absorbing trap of radius $R_t = 2\sigma$ at the center of the disk~\cite{vot_first-encounter_2022}. 

The exponential scaling of $T_{A+B \to AB}$ with $\e_{AB}^+$ is also verified for sufficiently large values of $\e_{AB}^+$ (Fig.~S\ref{fig:S1}B). However, while $T_d$ is a good approximation for two randomly placed particles to first diffuse towards each other, it does not adequately account for the succession of unbinding and rebinding events.   Indeed, a particle is more likely to (re)bind a target when it starts in its vicinity. Therefore, we estimate with the MD simulations the mean  time required for rebinding, $T_{d'}$, as a function of $L^2$. We then use $T_{d'}$ as a substitute for $T_d$. 

To describe the mean reaction time on a (auto)catalyst, we replace $T_{d'}$ by $T_c$, the average time for the substrates to encounter when constrained to remain bound to the (auto)catalyst. As shown in Fig.~S\ref{fig:S1}C, we verify that $T_{C(A+B) \to C(AB)}\approx T_{c}e^{\epsilon_{AB}^+}$ for sufficiently large $\e_{AB}^+$, consistent with Arrhenius law. 

\begin{figure}[t]
    \centering
    \includegraphics[width=.75\linewidth]{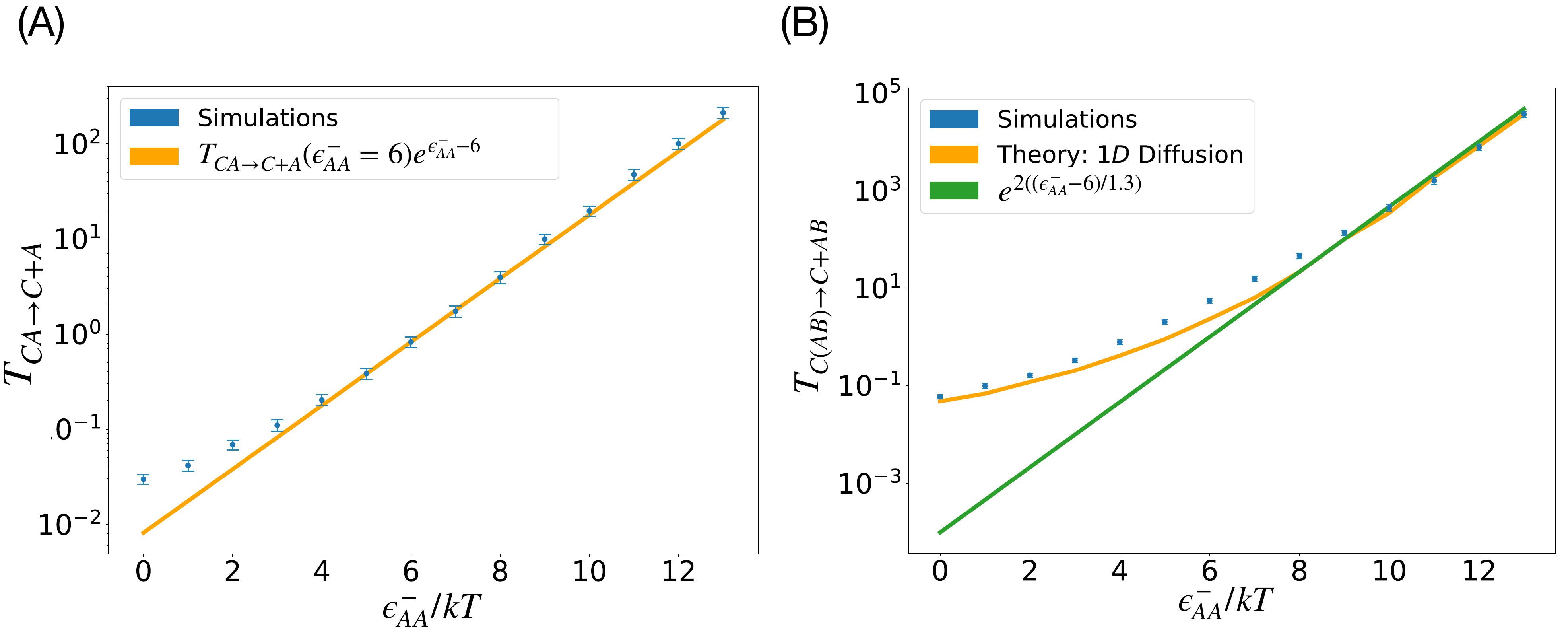} 
    \caption{\small Release of one substrate and of the product from the (auto)catalyst \textbf{(A)} Mean release time of a substrate $T_{CA \to C+A}$ as a function of the interaction energy $\e_{AA}^-$. We verify Arrhenius law for sufficiently large $\e_{AA}^-$, specifically when $\e_{AA}^-\gtrsim 6$ $k_BT$. \textbf{(B)} Mean release time of a product $T_{C(AB) \to C+AB}$ as a function of the interaction energy $\e_{AA}^-$. For low values of $\e_{AA}^-/k_BT$, the mean release time can be computed by considering an intermediate state \(C {\cdot} AB\) from which the likelihood -- calculated as a $1D$ diffusion problem~\cite{redner_guide_2001} -- that a detached particle reattaches before the other particle detaches is high. For large values of $\e_{AA}^-/k_BT$, the mean time for product release is well approximated by the time for two particles to detach simultaneously, \(T_{C(AB) \to C + AB} \approx (T_{A\cdot C +B\to A+B+C})^2\).} \label{fig:S2}
\end{figure}

To describe the mean time for a product $AB$ to dissociate from a (auto)catalyst, we first verify that the release of a single monomer scales exponentially with the interaction strength $\e_{AA}^-$ (Fig.~S\ref{fig:S2}A), consistent again with Arrhenius equation. We model product release as a mean first passage time from the state $C(AB)$ to $C + AB$ via a state $C{\cdot}AB$ in which only one particle of the dimer interacts with the (auto)catalyst. In state \(C{\cdot}AB\), the likelihood that a detached particle reattaches before the other particle detaches is high, due to its close proximity with the (auto)catalyst. This likelihood can be calculated as a $1D$ diffusion problem~\cite{redner_guide_2001} (Fig.~S\ref{fig:S2}B, orange line). At high interaction strength $\e_{AA}^-$, the likelihood that a detached particle reattaches before the other particle detaches is so high that we can approximate product release by a single step $(AB)(AB)\to 2AB$ with rate $e^{-2\e_{AA}^-}$, corresponding to the simultaneous breaking of two bonds of interaction strength $\e_{AA}^-$ (Fig.~S\ref{fig:S2}B, green line).

\begin{figure}[t!]
    \centering
    \includegraphics[width=.75\linewidth]{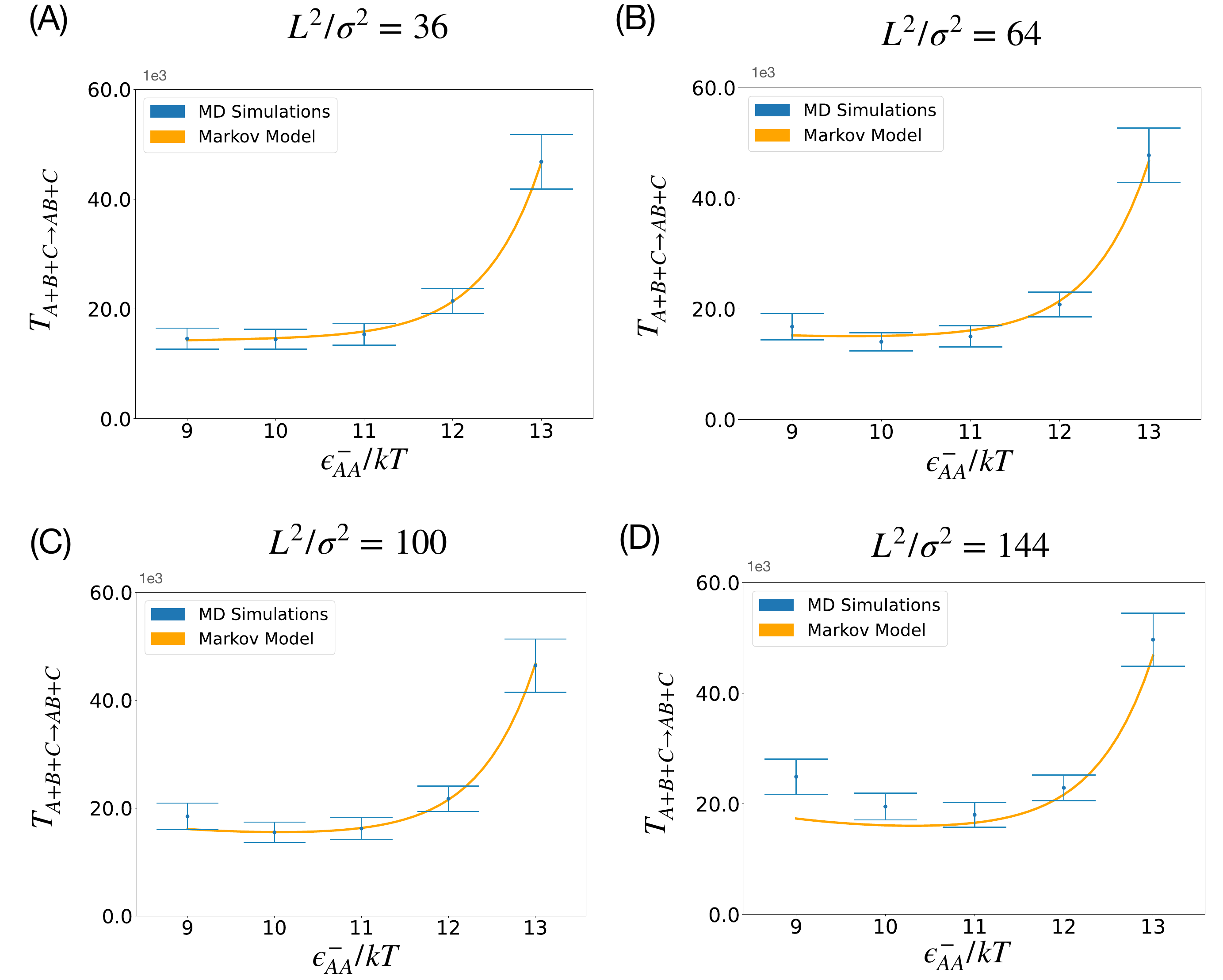} 
    \caption{\small Mean time to form a product \(AB\) in the presence of an (auto)catalyst \(C\) for different reaction areas \(L^2\) and interaction strengths \(\epsilon^-_{AA}\). These results demonstrate how the Markov model that we have derived fits quantitatively the MD simulations.}\label{fig:S3}
\end{figure}

Overall, from the mean times for the elementary steps of diffusion, dimerization and release, we define a Markov model, whose fit with the MD simulations is represented in Fig.~S\ref{fig:S3}. For low energy barriers ($\e_{AB}^+ < 6$, $\e_{AA}^- < 6$), the mean time of those processes are directly taken from the MD simulations. For higher energy barriers, when Arrhenius equation applies, we report the rates in the second column of Table~\ref{table:model}. 

To simplify the presentation, we analyze in the main text and in Sec.~\ref{sec:limitingbarrierSup} below a simplified model that ignores some of the pre-factors to retain only the dependence on the parameters. The reaction rates of this simplified model are listed in the third column of Table~\ref{table:model}. In Sec.~\ref{sec:full}, we verify that our results are largely unaffected by this simplification.

\begin{table}[ht]
\centering
\begin{tabular}{|l|l|l|}
\hline
Reactions & Comprehensive Markov model ($\e_{AB}^+, \e_{AA}^- > 6$) & Simplified Markov model \\ \hline
$A+B \to AB$ & $e^{\e_{AB}^+-6}/T_{d'}$  & $e^{\e_{AB}^+}/L^2$ \\ \hline
$A+B+C \to CA + B$ & $2/T_{d'}$ & $2/L^2$ \\ \hline
$CA +B \to A+B+C$ & $0.9e^{\e_{AA}^- - 6}$ & $e^{\e_{AA}^-}$ \\ \hline
$CA +B \to C(A+B)$ & $1/T_{d'}$ & $1/L^2$ \\ \hline
$C(A+B) \to CA + B$ & $1.8e^{\e_{AA}^- -6}$ & $2e^{\e_{AA}^-}$ \\ \hline
$C(A+B) \to C(AB)$ & $T_{c}e^{\e_{AB}^+-6} \approx 266e^{\e_{AB}^+ -6}$ & $e^{\epsilon_{AB}^+}$ \\ \hline
$C(AB) \to C+AB$ & $0.81e^{2(\e_{AA}^- - 6)/1.3}$ & $e^{2\epsilon_{AA}^-}$ \\ \hline
\end{tabular}
\caption{{\small Reactions and corresponding reactions rate of the comprehensive Markov model that fits the MD simulations, or its simplified version used in the main text. For the comprehensive Markov model, rate values are only shown for $\e_{AB}^+, \e_{AA}^- > 6$, where Arrhenius equation applies (otherwise, the rates are directly taken from the MD simulations).}} \label{table:model}
\end{table}

\subsection{Extension to anisotropic particles}

\begin{figure}[t!]
\centering
\includegraphics[width=0.5\linewidth]{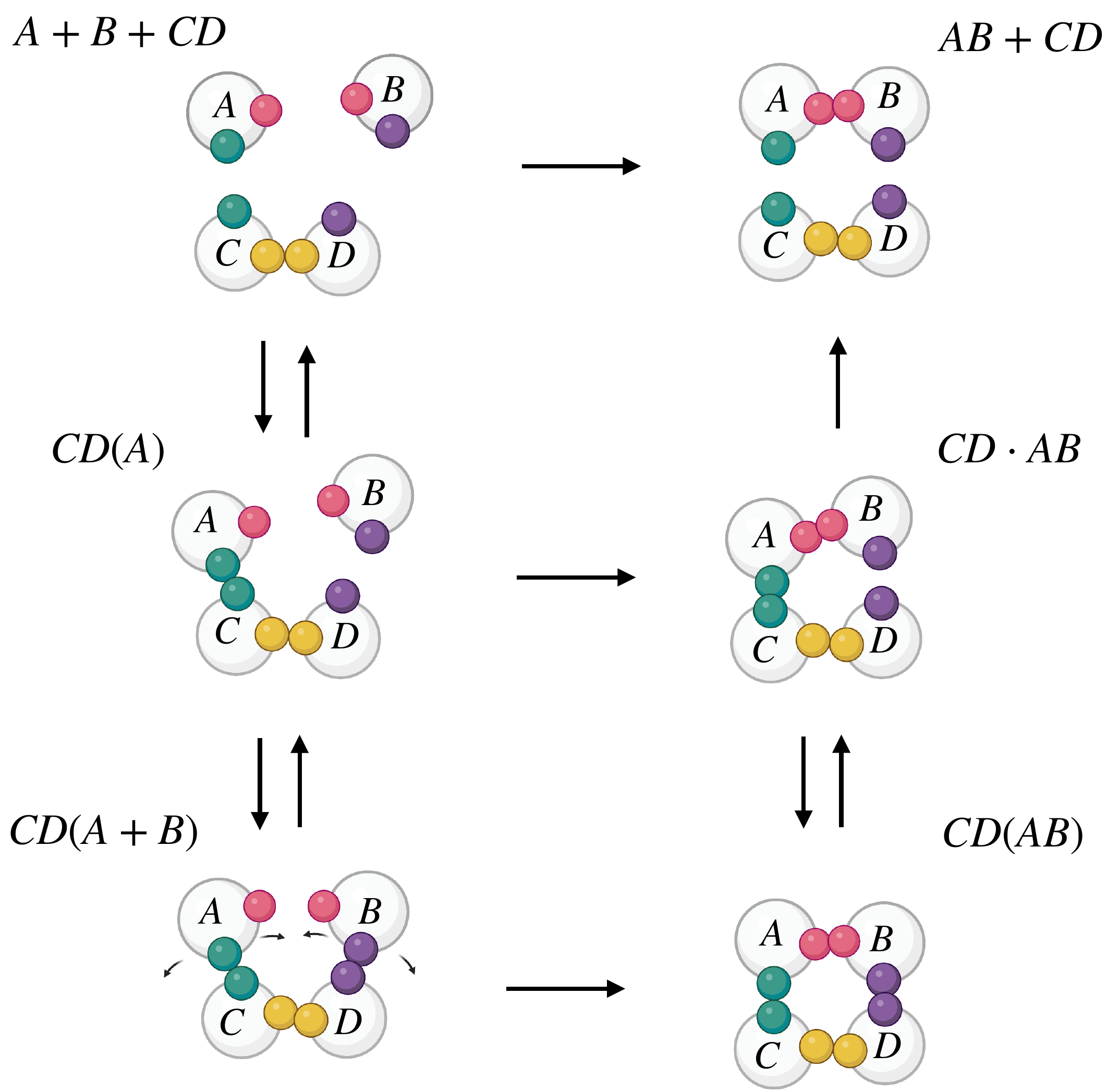}
\caption{\small Cross-catalysis with anisotropic (patchy) particles. Patches of the same colors can interact one with another. The decomposition in $6$ state is the same as the one discussed in the case of isotropic particles, with the difference that the autocatalyst $AB$ is now a cross-catalyst $CD$. } \label{fig:S4}
\end{figure}

Here we show how the analysis can be extended to anisotropic particles. For simplicity, we illustrate an example of cross-catalysis, where $CD$ catalyzes $A+B \to AB$ and $AB$ catalyzes $C+D \to CD$, as represented in Fig.~S\ref{fig:S4}. There are two main distinctions compared to the case with isotropic particles.

\begin{figure}[h!]
\centering
\includegraphics[width=.75\linewidth]{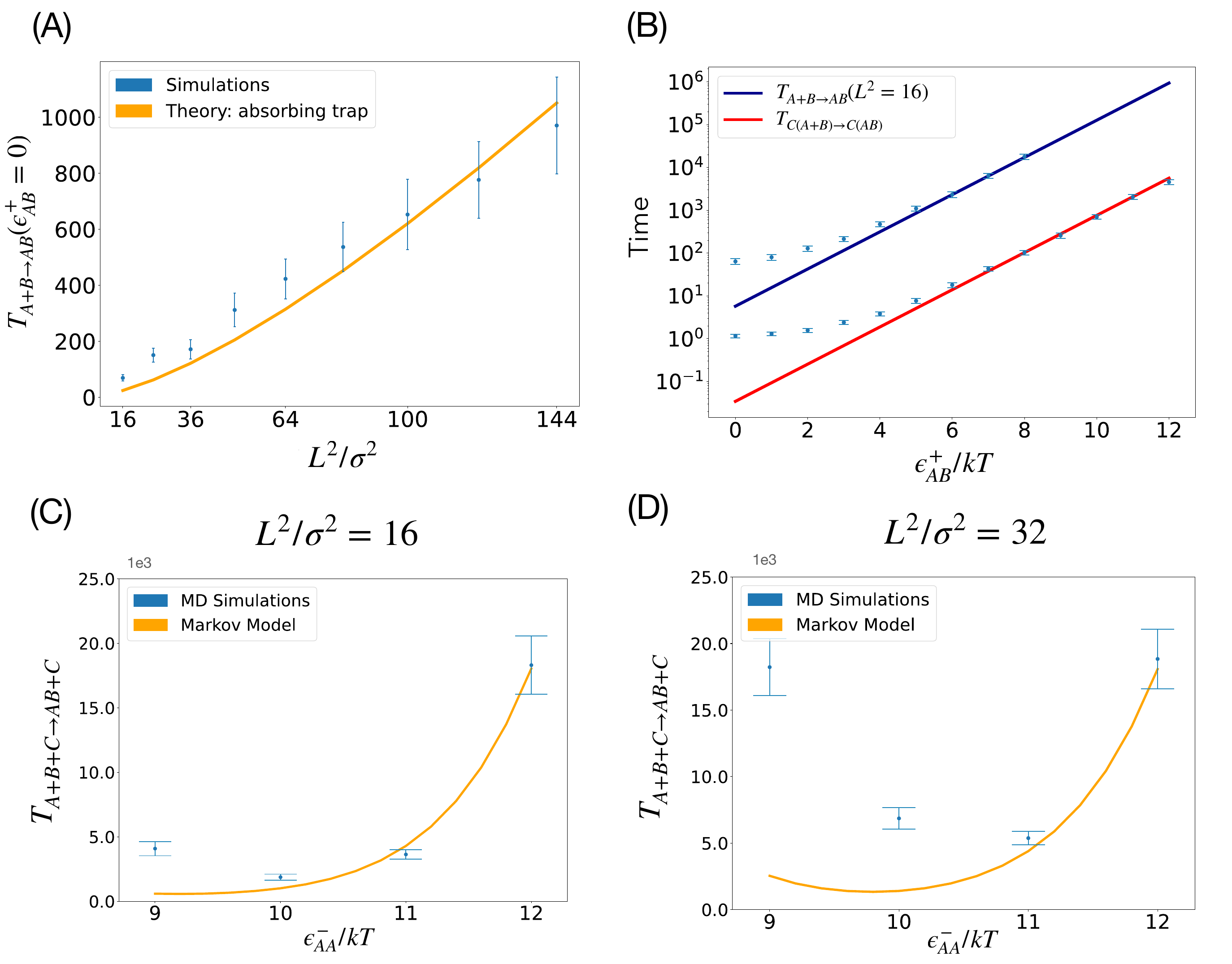}
\caption{\small MD simulations and Markov models with anisotropic particles. \textbf{(A)} Mean time required for two particles, initially positioned randomly, to diffuse towards each other and interact through their patches. The analysis reveals that this time is approximately \(1/10\) of the diffusion time for isotropic particles. \textbf{(B)} Representation of the dimerization reaction in the bulk (purple line) or on the catalyst (red line), plotted against the interaction barrier \(\e_{AB}^+\) showing how the dimerization reaction on the autocatalyst occurs swiftly, since the particles are suitably oriented for interaction. \textbf{(C) - (D)} Mean times to form a product with a catalyst in the reaction vessel, as computed by both MD simulations and a Markov model for anisotropic particles.}\label{fig:S5}
\end{figure}

First, diffusion is effectively longer, as it includes a rotational diffusion to orient particles with respect to each other. Formally, if the patches cover a fraction $r$ of the particles, only a fraction $r^2$ of all encounter events lead to an actual interaction, resulting in $T_d^{\rm patch} = T_d r^{-2}$. However, for patches of sizes $0.1\sigma$, MD simulations yield $r^{-2} \approx 10$, in contrast to the anticipated $r^{-2} \approx 100$ (Fig.~S\ref{fig:S5}A). This reduced entropic barrier arises from the particles not diffusing away once they come into proximity with each other~\cite{berg_orientation_1985}, a phenomenon of the same nature than the previously discussed difference between $T_d$ and $T_d'$. This effect occurs in both 2D and 3D environments and has been extensively studied ~\cite{ berg_orientation_1985,olc_kinetics_1973, shoup_diffusion-controlled_1981, eun_effects_2020}.

Second, patches can position the substrates nearer to their transition state, leading to a shorter dimerization on the catalyst, represented by $T_c$ (Fig.~S\ref{fig:S5}B). As previously, $T_c$ scales with barrier but can be significantly smaller than $T_d$ even when the later is estimated in confined areas.

\section{Limiting barriers} \label{sec:limitingbarrierSup}

\subsection{Limiting barrier in the absence of product}

Here we detail the derivation of the limiting barrier of the catalytic cycle for the dimerization reaction as a function of the concentration of the monomers, $[A] = [B]$, the interaction strength between particles of the same types, $\e_{AA}^-$, and the interaction barrier $\e_{AB}^+$. This derivation is based on the simplified Markov model of Table~\ref{table:model}.

In the absence of product, we obtain the kinetic barriers for the catalytic cycle are given by Eqs. (6) and (7)  in the main text:
\begin{align*}
    G_1^{\ddagger} -G_1 &= -\ln[A] - \ln2 & G_{2}^{\ddagger} - G_{1} &= -2\ln[A] - \e_{AA}^- - \ln2 \\
    G_2^{\ddagger} -G_2 &= -\ln[A] & G_{3}^{\ddagger} - G_{1} &= -2\ln[A] - 2\e_{AA}^- + \e_{AB}^+ \\
    G_3^{\ddagger} -G_3 &= \e_{AB}^+ & G_{3}^{\ddagger} - G_{2} &= -\ln[A] - \e_{AA}^- + \e_{AB}^++ \ln2 \\
    G_4^{\ddagger} -G_4 &= 2\e_{AA}^-.
\end{align*}
The limiting barrier is the largest of these barriers. It depends on the parameters and can be determined either analytically or numerically. The limiting barrier $G_j^\ddagger-G_i$ is said to be direct if $i=j$ and indirect if $i<j$. Overall, we find the following:

At high values of $-\ln[A]$, corresponding to low substrate concentrations, the dominating barriers are indirect barriers, namely $G_{2}^{\ddagger} - G_{1}$, which represents the diffusion of two substrate molecules to the catalyst, and $G_{3}^{\ddagger} - G_{1}$, the diffusion of the substrate followed by a subsequent dimerization reaction. Specifically, $G_{3}^{\ddagger} - G_{1}$ surpasses $G_{2}^{\ddagger} - G_{1}$ when the duration of the chemical step significantly exceeds the substrate release time, $\e_{AB}^+ > \e_{AA}^- - \ln2$. The threshold for $a$ to be considered ``high'' is determined by the conditions under which these dominating barriers are indeed larger than the other barriers, leading to $-\ln[A] > (3\e_{AA}^-)/2 + 1/(2\ln2)$ and $a > \e_{AB}^+ + 2\ln2$. 

When $\e_{AB}^+$ takes high relative values, the limiting barriers are either direct or indirect barriers associated with the chemical step, $G_3^{\ddagger} -G_3$, $G_{3}^{\ddagger} - G_{1}$, or $G_{3}^{\ddagger} - G_{2}$. In particular, the direct barrier, $G_3^{\ddagger} -G_3$, dominates when the duration of the chemical step exceeds that of the release, $\e_{AB}^+ > 2\e_{AA}^-$. This dominance also requires that the interaction strength significantly exceeds the rate of diffusion, expressed as $\e_{AA}^- > -\ln[A] + \ln2$. Conversely, in scenarios where these conditions are not met, the indirect barriers $G_{3}^{\ddagger} - G_{1}$ or $G_{3}^{\ddagger} - G_{2}$ dominate.

Finally, when $\e_{AA}^-$ is sufficiently large, the barrier associated with product release, $G_4^{\ddagger} -G_4$, dominates. These results are summarized in Fig.~1B of the main text for $\e_{AB}^+=1$.

\subsection{Limiting barrier in the presence of product}

In the presence of the product, three new barriers arise,
\begin{equation*}
\begin{aligned}
    &G_1^{\ddagger} -G_0 = -\ln[A] +2\e_{AA}^- + \ln[AB] - \ln2\\
    &G_{2}^{\ddagger} - G_{0}= -2\ln[A] +\e_{AA}^- +\ln[AB] - \ln2\\
    &G_{3}^{\ddagger} - G_{0}= -2\ln[A] + \e_{AB}^+ +\ln[AB]\\
\end{aligned}
\end{equation*}
They are equivalently written as  $G_i^{\ddagger} -G_0 = G_i^{\ddagger} -G_1 +\ln[AB] + 2\e_{AA}^-$ for $i = 1, 2, 3$, thus indicating that these barriers dominate the respective barriers $G_i^{\ddagger} -G_1$ only when $2\e_{AA}^- >-\ln[AB]$, that is, for high concentration of product relative to the interaction strength. In such cases, $\Tc \approx \Ti$. Moreover, we find that $G_1^{\ddagger} -G_0$ and $G_2^{\ddagger} -G_0$ dominate over $G_4^{\ddagger} -G_4$ when $-\ln[A]> - \ln[AB] + \ln2$ and $-2\ln[A] > \e_{AA}^- -\ln[AB] + \ln2$, that is, when the concentration of substrate is lower than that of the product. These results are summarized in Fig.~4A of the main text.

\section{Growth laws}

\begin{figure}[t]
    \centering
    \includegraphics[width=1\linewidth]{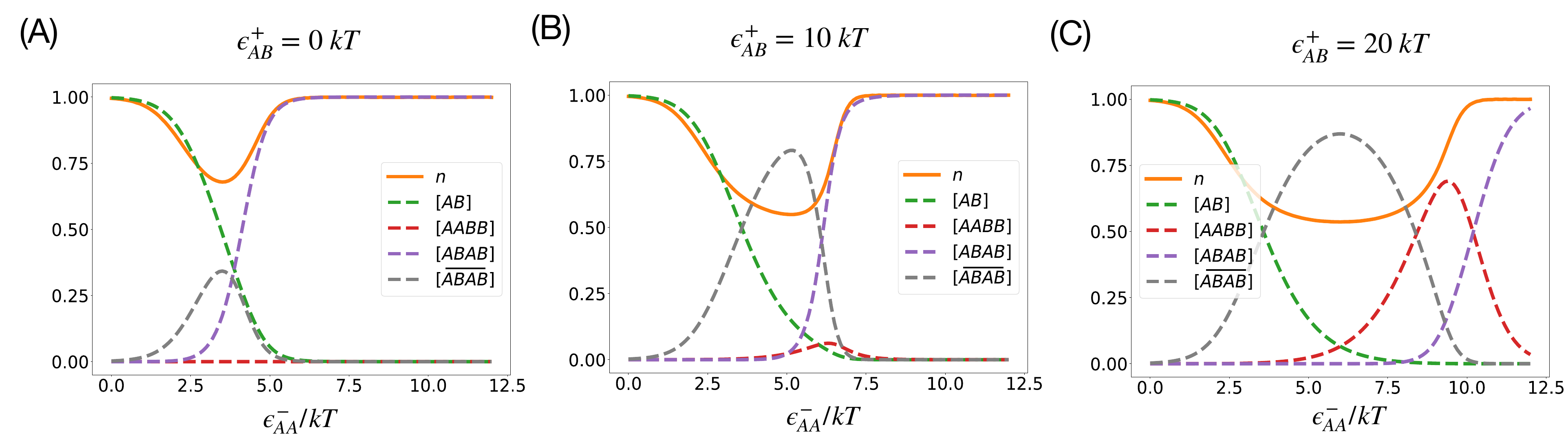} 
    \caption{\small Reaction order and relative species concentration as a function of the interaction strength $\e_{AA}^-$ for different values of the interaction barrier $\e_{AB}^+$. The reaction order is maximal ($n=1$) when the interaction strength is either weak or strong. At intermediary strength, however, $0.5<n<1$, due to the presence of product inhibition (high $\overline{[ABAB]})$. We verify that $[AB]\approx [AB]_{\rm tot}$ for weak interaction strength, and $[AB]\approx 0$ at large interaction strength.} \label{fig:S6}
\end{figure}

\subsection{Growth regimes}

Instead of a fixed concentration of free product $[AB]$ (fixed parameter $-\ln[AB]$), we may consider a fixed total concentration of product $[AB]_{\rm tot}$, including products that interact with catalysts. The difference is illustrated in Fig.~4 of the main text showing that the limiting barrier is changed only for high values of the interaction strength $\e_{AA}^-$. This is because for small $\e_{AA}^-$, most products are in free form, $[AB] \approx [AB]_{\rm tot}$, while for large $\e_{AA}^-$, most products are in complexes. 

More quantitatively, the regime of small $\e_{AA}^-$ requires $\e_{AA}^- < - \ln[AB]/2$ and  $\e_{AA}^- -\ln[AB] + \ln2<-2\ln[A]$. Because increasing $\e_{AA}^-$ affects the limiting barrier both directly (as per the formulas) and indirectly (through decreasing the concentration of $[AB]$), the regime of high $\e_{AA}^-$ is less easy to determine precisely in general. However, in conditions where the chemical step is much longer than all other processes ($\e_{AB}^+ \gg -\ln[A], 2\e_{AA}^-)$, von Kiedrowski showed that the reaction order $n$ can be estimated as~\cite{dugas_minimal_1993}
\begin{equation}
n = \frac{4K_2[AB]_{\rm tot} q^2}{8K_2cq^2 + (1+q)^2 - (1+q)\sqrt{8K_2cq^2 + (1+q)^2}},
\end{equation}
where $q = 1/(K_1[A]^2)$, $K_1=(k_1/k_{-1})^2 = e^{2\e_{AA}^-}$, and $K_2=(k_2/k_{-2})^2 = e^{2\e_{AA}^-}$. From this expression, it follows that exponential growth occurs when release is limiting, that is when $K_1[A]^2 \gg \sqrt{2K_2[AB]_{\rm tot}}$ and $K_1[A]^2 \gg 1$. We verify these different results in Fig.~S\ref{fig:S6} where the reaction order is computed numerically (as per Material and Methods) as a function of the interaction strength $\e_{AA}^-$ for different values of the reaction barrier $\e_{AB}^+$.\\

\begin{figure}[t]
    \centering
    \includegraphics[width=.85\linewidth]{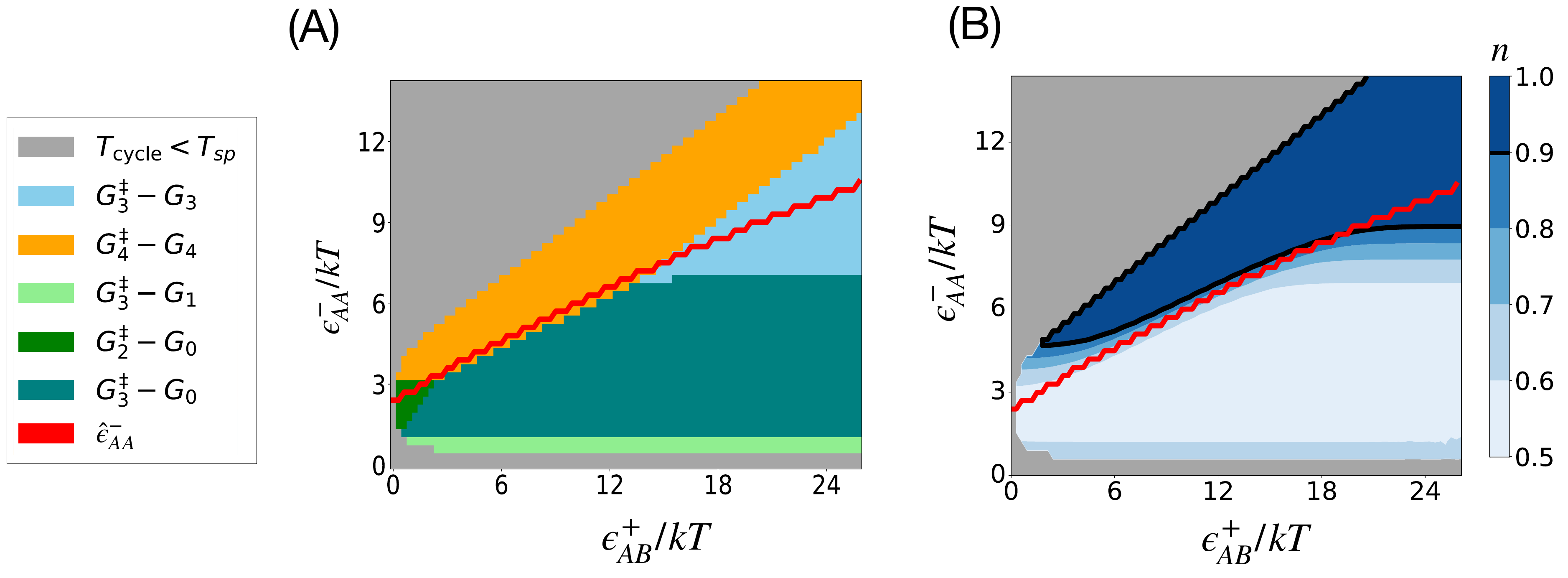} 
    \caption{\small Relationship between $k$ and $n$ as a function of the interaction barrier $\e_{AB}^+$. At low interaction barrier, the interaction strength $\e_{AA}^-$ that maximizes $k$ occurs for $n\approx0.5$. In this regime, the limiting barrier is $G_{2}^\ddagger - G_0$. As the interaction barrier increases, the optimal $k$ is reached for $n$ closer to $n=1$. At high interaction barriers, the limiting barrier is $G_{3}^\ddagger - G_3$, and $k$ and $n$ are not in trade-off anymore. Here $[A]=e^{-4}$ and $[AB]_{\rm tot} = e^{1}$.} \label{fig:S7}
\end{figure}

\subsection{Trade-off between the rate $k$ and the reaction order $n$}

Exponential growth ($n=1$) is achieved for either low or high interaction strength $\e_{AA}^-$ (Fig.~S\ref{fig:S6}), but maximal turnover ($k^{-1}$) is achieved for an intermediary interaction strength (Sabatier principle). This raises the question of a trade-off between $k$ and $n$. As shown in Fig.~S\ref{fig:S7}B, this trade-off is present only for some values of $\e_{AB}^+$. For small values of $\e_{AB}^+$, the optimal $k$ is achieved while $n=1/2$, implying a clear trade-off, but for larger values of $\e_{AB}^+$, the optimal $k$ is achieved while $n\lesssim 1$, implying a marginal trade-off, while for even larger values of $\e_{AB}^+$, the optimal $k$ is achieved while $n= 1$, implying no trade-off.

This is rationalized by recognizing that the limiting barrier defines different regimes as a function of $\e_{AB}^+$ (Fig.~S\ref{fig:S7}A). For small $\e_{AB}^+$, $k$ is optimized when $G_2^\ddagger - G_0$ is limiting, which is a barrier caused by product inhibition. For larger $\e_{AB}^+$, $G_3^\ddagger - G_0$ dominates over $G_2^\ddagger - G_0$. This barrier is also caused by product inhibition but does not depend on $\e_{AA}^-$. It is therefore minimized by minimizing $[AB]$, until the point where $G_3^\ddagger-G_0$ equates $G_4^\ddagger-G_4$, which depends on $\e_{AA}^-$. The optimal interaction strength is then found at the boundary between $G_3^\ddagger-G_0$ and $G_4^\ddagger-G_4$, corresponding to $n\lesssim 1$. For even larger $\e_{AB}^+$, $k$ is optimized when the limiting barrier becomes $G_3^\ddagger-G_3$, which is unrelated to product inhibition, leading to $n=1$.

\subsection{Exponential growth and limiting barriers with the comprehensive Markov model}\label{sec:full}

\begin{figure}[t]
    \centering
    \includegraphics[width=.85\linewidth]{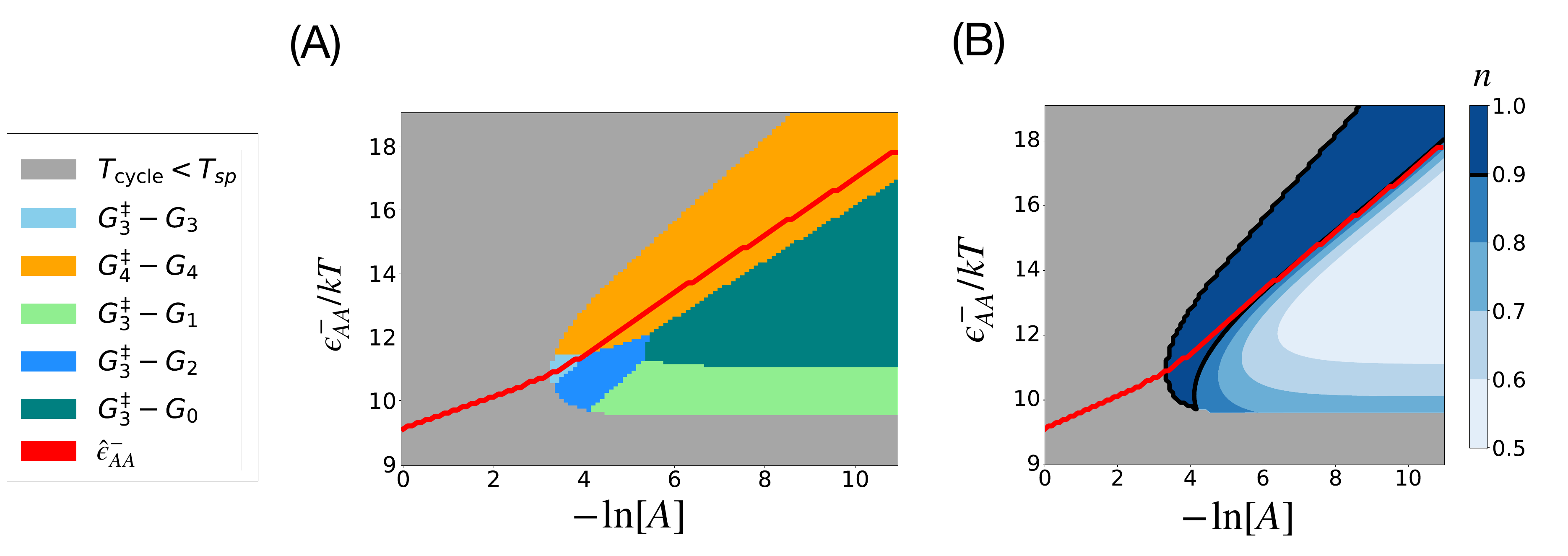} 
    \caption{\small Limiting barriers in the presence of products. \textbf{A.} Reaction order $n$, as computed from simulations of the ordinary differential equations describing the specific Markov model, when fixing the total concentration of autocatalyst $[AB]_{\rm tot}$. The results are comparable to Fig.~4C, although shifted from stronger interaction strengths (note the difference of scale for the y-axis). \textbf{B.} Limiting barriers when fixing the total concentration of autocatalyst to $[AB]_{\rm tot} = e^{-3}$. The results are comparable to those of Fig.~4B.} \label{fig:S8}
\end{figure}

To simplify the presentation, we analyze in the main text a Markov model that is a simplified version of the Markov model obtained from the MD simulations (supplementary section~\ref{sec:model}). In Fig.~S\ref{fig:S8}, we replicate Fig.~4B-C with the more comprehensive Markov model. The comparison between Fig.~S\ref{fig:S8} and Fig.~4B-C shows that the two models lead to very comparable results. The primary distinction is that a greater interaction strength $\e_{AA}^-$ is necessary for (auto)catalysis to take place. Crucially, for such strong interactions, pre-factors of order $1$ are negligible, and the simplified model is therefore justified.

